\documentclass{article}
\usepackage[utf8x]{inputenc}
\usepackage{natbib}
\setcitestyle{numbers}
\usepackage{graphicx}
\usepackage{amsmath}
\usepackage{float}
\usepackage[letterpaper, margin=1.25in]{geometry}
\setcitestyle{comma}
\setcitestyle{square}

\begin{document}
\begin{flushleft}
%\linenumbers

\textbf{Title:} A Testbed for Investigation of Selective Laser Melting at Elevated Atmospheric Pressure

\textbf{Authors:} David A.\,Griggs\textsuperscript{1}, Jonathan S.\,Gibbs\textsuperscript{1,3}, Stuart P.\,Baker\textsuperscript{1,2,4}, Ryan W.\,Penny\textsuperscript{1}, Martin C.\,Feldmann\textsuperscript{1}, A.\,John Hart\textsuperscript{1}

\textbf{Affiliations:} 

1. Department of Mechanical Engineering, Massachusetts Institute of Technology, Cambridge, MA 02139 USA
                2. Department of Electrical Engineering and Computer Science, Massachusetts Institute of Technology, Cambridge, MA 02139 USA

3. Department of Naval Architecture and Ocean Engineering, U.S. Naval Academy, Annapolis, MD, USA

4. Materials and Manufacturing Directorate, Air Force Research Laboratory, Dayton, OH

\setlength{\parskip}{1em}

\textbf{Corresponding Authors:} David A.\,Griggs (griggs@alum.mit.edu), A.\,John Hart (ajhart@mit.edu)

\textbf{Abstract:} Metal additive manufacturing (AM) by laser powder bed fusion \mbox{(L-PBF)} builds upon fundamentals established in the field of laser welding which include the influence of gas and plume dynamics on weld depth and quality. \mbox{L-PBF} demands a thorough investigation of the complex thermophysical phenomena that occur where the laser interacts with the metal powder bed. In particular, melt pool turbulence and evaporation are influenced by the ambient gas chemistry and pressure. This paper presents the design and validation of high pressure laser melting (HPLM) testbed; this accommodates bare metal plate samples as well as manually-coated single powder layers, and operates at up to 300\,psig. The open architecture of this testbed allows for full control of all relevant laser parameters in addition to ambient gas pressure and gas flow over the build area. Representative melt tracks and rasters on bare plate and powder are examined in order to validate system performance, and preliminary analysis concludes that pressure has a significant impact on melt pool aspect ratio. The HPLM system thus enables careful study pressure effects on processing of common \mbox{L-PBF} materials, and can be applied in the future to materials that are challenging to process under ambient pressure, such as those with high vapor pressures.


\section{Introduction}
Laser powder bed fusion (L-PBF) is a category of additive manufacturing (AM) in which laser energy is focused onto a flat bed of powder to progressively fuse together horizontal cross-sections of a three-dimensional component. Figure \ref{fig:L-PBF_Def}a depicts the process: powder is spread across a build platform to form a thin layer; the laser is focused to the flat powder bed and directed by galvanometer mirrors in order to fuse an arbitrary shape in the powder; an inert gas atmosphere minimizes oxidation, and additionally the inert gas is swept across the build platform as a so-called ``gas knife'' to remove process byproducts from the chamber.

\begin{figure}[H]
    \centering
    \includegraphics[width=1.0\textwidth]{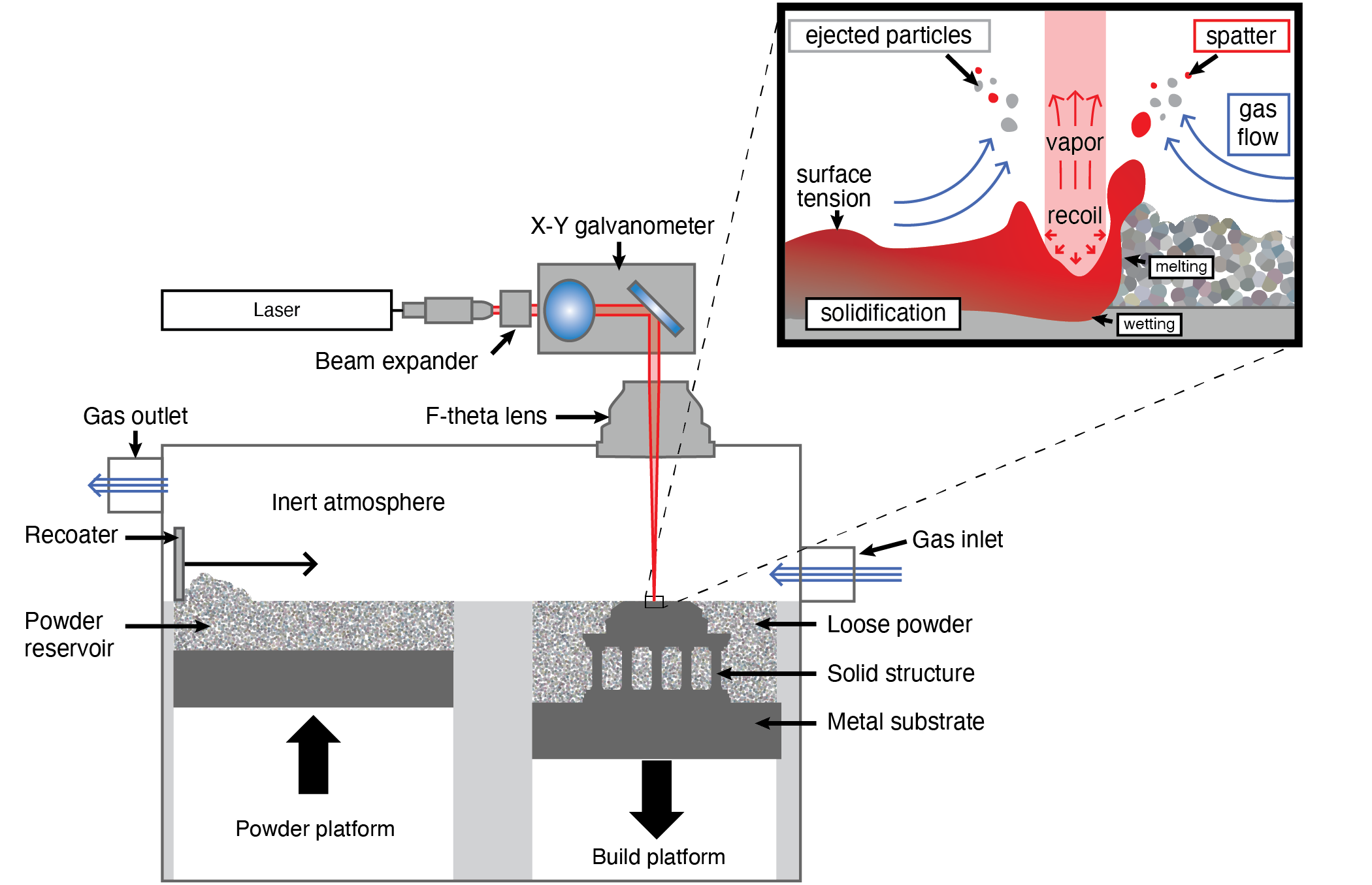}
    \caption[Schematic of a typical L-PBF machine and the mesoscale dynamics of the L-PBF process]{Schematic of a typical L-PBF machine and the mesoscale dynamics of the L-PBF process. Adapted from \cite{Meier2017ThermophysicalExperimentation}.}
    \label{fig:L-PBF_Def}
\end{figure}

In L-PBF, absorption of laser energy is governed by the powder material itself, as well as the prevalence of secondary reflections as impacted by powder layer properties and previously fused geometry \cite{Solana1997AMaterials,Gusarov2009Two-dimensionalMelting,Boley2017CalculationManufacturing,Trapp2017InManufacturing,Khairallah2020ControllingPrinting}. The resulting melt pool may or may not reach a steady state, subject to a number of competing forces as the laser progresses across the powder bed. The laser continues heating the melt pool while heat is convected, conducted, and radiated away from the laser spot \cite{Gusarov2009Two-dimensionalMelting,Khairallah2016LaserZones}, which establishes a temperature gradient. This gradient creates a corresponding surface tension gradient that induces Marangoni flow, which in most cases circulates from the laser spot where the surface tension is lowest, to the edges of the melt pool where surface tension is highest \cite{Sun2017PowderOverview}. The velocity, and thus turbulence, of this flow grows with greater energy input to a point at which the melt pool becomes unstable \cite{Pei2017NumericalPowder} and the intense internal flows may cause voids in the final part \cite{ScipioniBertoli2017OnMelting}.

In addition to liquid flow, gas-liquid interaction plays a significant role in melt pool behavior and resulting L-PBF part quality. Khairallah et al.\,\cite{Khairallah2016LaserZones} describe in detail how evaporation from the molten metal surface generates a recoil pressure above the melt pool, which presses down into the melt pool with sufficient force at high energy densities to create keyhole porosity. This injects gas into the melt pool and forces liquid to the sides of the depression with enough force to project spatter upward and away from the laser spot at high speeds. Matthews et al.\,\cite{Matthews2016DenudationProcesses} observed the denudation of powder with high speed imaging at pressures from 0.5\,Torr to 1\,bar. It was concluded that denudation of powder near the laser spot is strongly influenced by ambient pressure, and that at pressures near 1\,bar, denudation decreases with pressure. Vaporization due to thermal gradients near the center of the laser spot poses an additional complication when some elements that comprise the working material vaporize more readily than others, resulting in a change in final part composition relative to the feedstock \cite{Wei2015InfluenceComponents,Zhang2020ElementMelting}. 

In summary, an ideal melt pool receives enough energy to maintain a single, continuous track which reaches the previously-solidified layer below to fuse the layers together, yet not so much energy that the melt pool becomes unstable, resulting in non-uniformity and/or embedded porosity. Much work has gone into mapping the parameter space for various materials in order to establish optimal processing conditions, sometimes called a process window, for high quality L-PBF parts. Recent work focusing on laser welding and L-PBF in sub-atmospheric conditions suggests that ambient gas pressure may be used to expand such process windows. 

Specifically, weld penetration increases and melt pool width decreases with decreasing pressure below 1\,bar \cite{Katayama2011DevelopmentVacuum, Jiang2017EffectWelding,Li2018ExperimentalPressure,Jiang2019ComparisonSub-atmosphere,Sokolov2015ReducedSteel}. For example, Pang et al.\,\cite{Pang20153DEffect} developed a 3D multiphase model of a laser weld keyhole and the resultant vapor plume, and found that adding ambient pressure into the simulation resulted in much lower (and more accurate) gas plume velocity estimates. Additionally, Masmoudi\,\cite{Masmoudi2015InvestigationProcess} presented a 3D numerical model to investigate laser-material-atmosphere interactions at low pressures, and, based on trends observed, proposed that elevated ambient gas pressure may decrease vapor convection near the melt pool, and reduce overall vaporization.

To our knowledge, there is only one report in the literature of L-PBF at pressures greater than 1 bar. Bidare et al.\,\cite{Bidare2017AnMeasurements} describe the design of an open architecture L-PBF testbed which was later outfitted with a vacuum chamber to observe L-PBF of a single powder layer at pressures between 10\,$\mu$bar and 5\,bar \cite{Bidare2018LaserPressures,Bidare2018LaserAtmospheres}. Using Schlieren imaging, they observe the plume characteristic to L-PBF and report that increased ambient pressure slows its speed and increases its temperature. They note that increasing ambient pressure significantly reduces denudation of surrounding powder as the laser progresses across the powder bed. They also find that melt pool depth is decreased with increasing ambient pressure up to 5 atm. Bidare et al.\,also suggest that elevated pressure results in more aggressive spatter and a decrease in laser energy reaching the working material. However, it is not clear that an adequate gas flow over the build surface was maintained to fully decouple the effects of elevated pressure from those of the plume obscuring and scattering laser energy, nor was the scaling of melt penetration quantified in detail.

Modifying an existing commercial L-PBF machine to achieve pressures significantly above one atmosphere would be challenging, and perhaps impossible given the gas flow and sealing requirements. Here, to investigate elevated pressure effects on L-PBF, we present a custom-built high pressure testbed. The system’s open architecture gives complete control over all relevant process variables, including laser power, scan speed, scan strategy, spot size, inert gas flow over the print area, and ambient pressure. In particular, the gas flow over the build area is simulated and validated, and the system is calibrated to ensure precise measurement of the laser spot size, and uniformity of exposure. Studies of melt track dimensions versus laser power and speed, and ambient pressure validate the system's baseline performance, and present preliminary insights as to the influence of elevated pressure on the L-PBF process window.

\section{Description of the HLPM Testbed} \label{sec:hardware}

\subsection{Chamber and build platform}

The high-pressure laser melting (HPLM) system was constructed upon a custom chamber (Parr Instrument Company) with a maximum operating pressure of 1900\,psig, and a cylindrical 2.5\,inch (diameter) by 6.5\,inch (length) bore. The chamber features an oblong window flange, a removable head with four ¼'' NPT ports on one end, and a single ¼'' NPT port on the opposing end. Sapphire was chosen as the pressure window material owing to its high fracture toughness and high transmission rate of wavelengths required for near- and mid-wave infrared (NIR, MWIR) optical process monitoring. A custom-designed 3D printed nozzle is attached internally to one of the four inlet ports, creating a gas ``knife'' which is designed to sweep vapor and byproducts away from the build area. Due to the short distance between the window and the build surface, a fused quartz microscope slide (Fisher Scientific) is suspended between the build area and the sapphire window by a thin frame resting on the build platform. This slide serves as a sacrificial shield to block spatter particles that may otherwise penetrate the gas knife and impact the window \cite{Zhang2020SimulationFusion}. The resulting full print area through the pressure window and microscope slide is approximately 10\,$\times$\,60\,mm, and the longer dimension is parallel to the gas flow.

\begin{figure}%[H]
    \centering
    \includegraphics[width=1.0\textwidth]{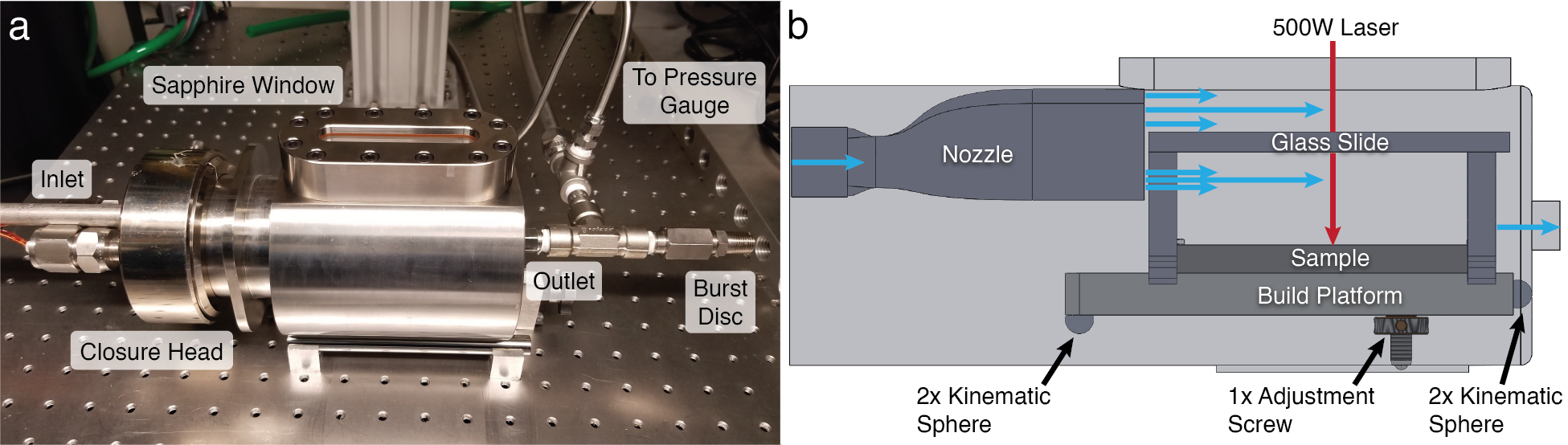}
    \caption{HLPM testbed: (a) the pressure chamber containing the build platform; (b) schematic of the internal architecture, noting the kinematic constraint of the platform assembly, and the placement of the gas nozzle with respect to the build platform, protective glass slide, and window. The chamber bore is cylindrical, 2.5\,inches (diameter) by 6.5\,inches (length).}
    \label{fig:chamber}
\end{figure}

Alignment between the laser’s focal plane and the powder bed is crucial for achieving uniform quality across the entire build area; misalignment will cause a change in effective laser spot size. To this end, a kinematic positioning system, displayed in Figure \ref{fig:chamber}, was devised to exactly constrain the build platform while giving precise control of alignment with the focal plane. Two spheres contact the cylindrical bore of the chamber. One adjustable ball-end screw contacts a small groove cut into the floor of the chamber, opposite the window flange above, to center the platform and allow for fine adjustment of the platform’s tilt to match the laser focal plane. Finally, two spheres contact the back wall of the chamber’s interior, fully constraining the platform in both rotation and translation. A sample plate (i.e., the substrate upon which L-PBF is performed) is bolted to the build platform, and the glass slide and its holder are placed atop the platform. The build platform, sample, and slide can thus be easily removed and replaced with high repeatability.

\subsection{Inert gas flow}

Achieving high quality L-PBF requires a steady, uniform flow of inert gas across the build area \cite{Ferrar2012GasPerformance,Philo2018AProcess,Reijonen2020OnManufacturing}. Regions of low speed flow below a certain threshold must be avoided, as even momentary local stagnation can result in defects due to inadequate removal of the melt-induced vapor plume and spatter. It has been recommended that a minimum gas velocity of 1\,-\,2\,m/s over the powder bed is required to obtain optimal L-PBF quality \cite{Reijonen2020OnManufacturing,Shen2020InfluenceFusion,Saunders2017GoneLinkedIn}. Although an even faster flow may be beneficial \cite{Ferrar2012GasPerformance,Ladewig2016InfluenceProcess,Anwar2017SelectiveStrength,Philo2018AProcess,Reijonen2020OnManufacturing}, at a certain velocity the gas flow will begin to disturb the powder bed below, which is undesirable. To this end, Shen et al.\,\cite{Shen2020InfluenceFusion}, building upon earlier work by Kalman et al.\,\cite{Kalman2005PickupParticles}, modeled and validated the upper bound velocity that will not disturb the powder bed. Therefore, a first step in our design process was to determine a target gas velocity for flow within the HPLM chamber. Following the above guidance, this was established to be  3.1\,m/s, for 15\,-\,45\,$\mu$m stainless steel powder (SS316L). 

One unique challenge arises as a result of the small size of the pressure chamber. Some amount of local recirculation is common in L-PBF, as the gas flow spreads and is not always fully captured by the exit port. Recirculating gas in the chamber may carry with it soot and particles from the melt pool, which may cross the path of the laser, deposit on the laser window, or alight on the powder bed \cite{Zhang2020SimulationFusion,Philo2018AProcess,Chen2018OptimizationChamber}. Most commercial L-PBF printers use build chambers with several inches or more of vertical space between the build plate and the laser window above, and ample lateral margin around the build area. In our much smaller chamber, any recirculation near the window is much more likely to deposit soot onto the window surface, which may attenuate and scatter laser energy and/or cause thermal damage to the window itself. 

Toward preventing recirculation, Chen\,\cite{Chen2018OptimizationChamber} and Saunders\,\cite{Saunders2017GoneLinkedIn} both suggest a secondary flow be established with a second nozzle near the top of the chamber. This can be seen in some commercial L-PBF machines, usually positioned near the laser optical window at the top of the chamber. In the present study, a representation of the chamber, build platform, and a simple nozzle were modeled using COMSOL Multiphysics software, and various strategies to prevent recirculation were modeled, including baffles, guide vanes, tapering (funneling) the flow path, and enlarging the outlet port. Each of these approaches merely shifted the location at which the gas stream splits and recirculates. However, the aforementioned secondary nozzle approach was found to successfully prevent a low pressure region from forming and causing recirculation near the HPLM chamber window. This solution is illustrated in parts a and b of Figure \ref{fig:CFD}. The final design was detailed in CAD (Solidworks) and brought into COMSOL via an integration module (Solidworks LiveLink). The critical dimensions were defined as the nozzle heights and widths (W1, H1, W2, H2), proportional entrance area at the split plane ($H3/H4$), and lengths (L1, L2). These were parametrically varied in the model, and design variations were judged on the velocity achieved above the entire print area, subject to the requirement for no recirculation near the laser window and the desire to minimize overall gas usage (max 15\,NLPM). The simulated flow field through the final nozzle design and chamber is shown in Figure \ref{fig:CFD}c. A physical model was printed with a Form2 3D printer (Formlabs) and attached to the gas flow system as shown in Figure \ref{fig:CFD}d. Fan (GM816, Amgaze) and hot wire (405i, Testo) anemometers were used to measure gas speeds at distances up to three inches away from both (top and bottom) nozzle outlets. It was found that with an input flow of 12\,-\,15\,NLPM, the final nozzle design produces gas velocities greater than $1.4\,m/s$ above the entire print area under ambient conditions.

\begin{figure}[H]
    \centering
    \includegraphics[width=1.0\textwidth]{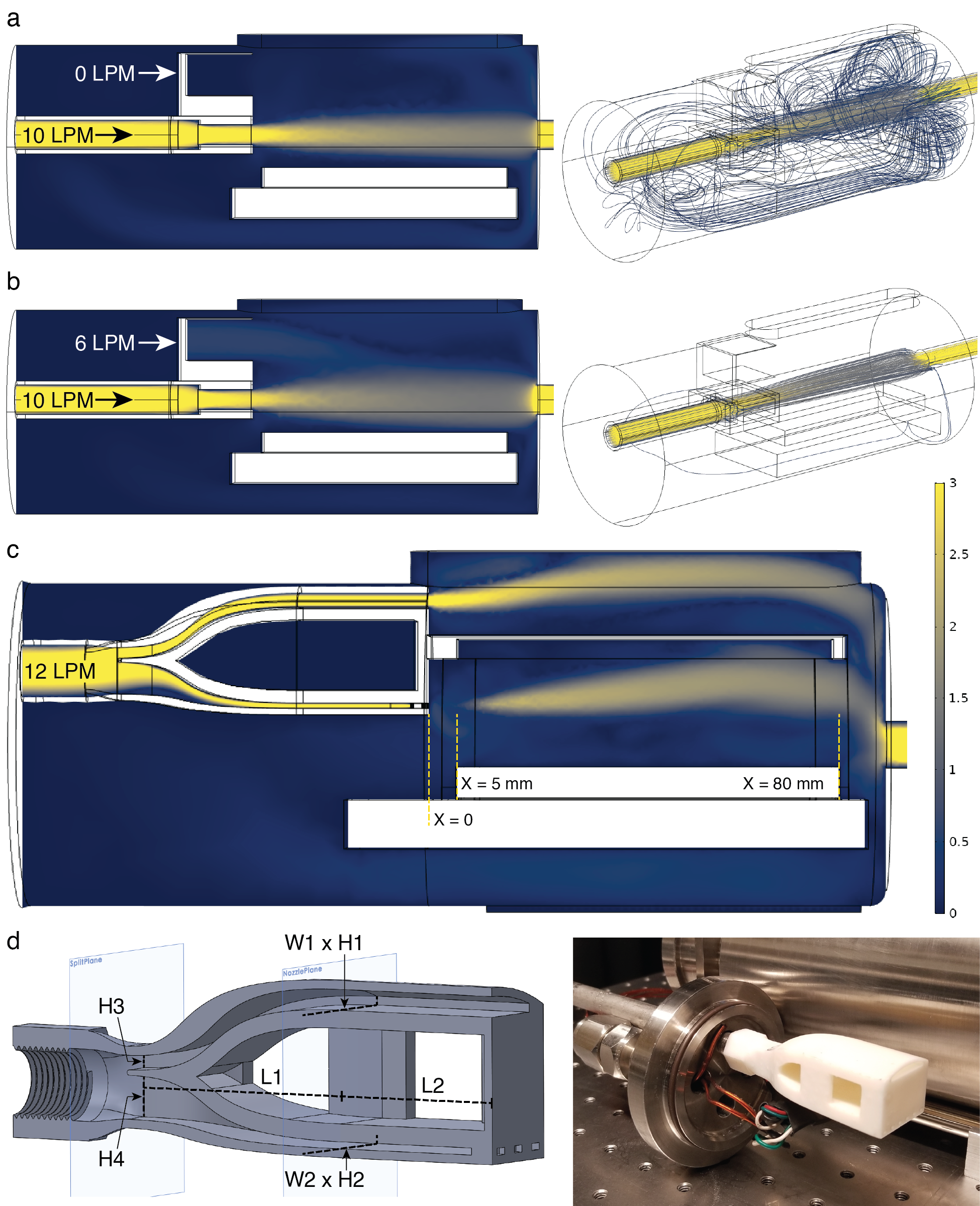}
    \caption{CFD analysis of gas flow within the HPLM chamber. Gas velocity along the central vertical plane of the chamber (a) without and (b) with a secondary nozzle. (c) Final design, for which the simulation predicts a gas velocity exceeding 1.5\,m/s at all points above the print area with an input flow of 12\,NLPM. All simulations shown are at 0\,psig. (d) 3D section view of the final nozzle design, and an image of a 3D printed nozzle coupled to the gas flow system.}
    \label{fig:CFD}
\end{figure}

Yet, as a certain volumetric gas flow is necessary to remove vapor and spatter from above the melt pool, it is necessary to use a higher mass flow with increasing pressure above 0\,psig. Additional COMSOL simulations showed that steady state gas flow in the chamber at pressures up to 1000\,psig remains qualitatively similar to that at atmospheric pressure. However, as pressure is increased in the chamber, argon density also increases nearly proportionately. This denser gas moving at the same velocity exerts more force on the powder particles, which effectively lowers the maximum gas velocity allowed before powder is disturbed. 

Additionally, the refractive index of argon is known to increase with density according to the Lorentz-Lorenz equation: 
\begin{equation}
    \frac{(n^2-1)}{(n^2+2)}=4\pi/3\cdot\rho\cdot\alpha
    \label{eq:lorenz}
\end{equation}
where n is refractive index, $\rho$ is the density $(mol/cm^3)$, and $\alpha$ is the polarizability of argon at atmospheric pressure $(cm^3/mol)$. Using a ray-tracing software (OpticStudio 16.5, Zemax), we found that the change in refractive index imparts a vertical shift in the location of the beam waist, effectively changing the spot size of the laser at a chosen focal plane. At 300\,psig the argon increases the spot size by approximately 16\,$\mu$m. This shift was not compensated for in our interpretation of pressure-dependent results, but we expect it accounts for a portion of the observed influence of pressure on apparent melt track dimensions under identical (input) laser parameters.

\subsection{Gas delivery system}

A flow-through system was devised using a standard Ultra High Purity (UHP) argon bottle with a regulator to supply inert gas to the chamber at pressures as high as the rated pressure of the chamber. The system diagram is shown in Figure \ref{fig:GasSystem}a. Needle valves upstream and downstream of the chamber allow the user to set both the pressure and flow of the argon stream within the operating limits of the system. A large filter rated to remove 95\% of 10\,nm diameter particles (United Filtration Systems, SLH818) protects equipment further downstream while adding minimal pressure drop to the system. A flowmeter calibrated for argon (Sierra Instruments, M100H1), reads out the mass flow through the system at a range of 10\,-\,1100\,NLPM with 1\%FS accuracy, which accommodates performance at pressures up to 1200\,psig. A smaller flowmeter (Sierra Instruments, M100L) can be swapped in to measure up to 50\,NLPM with 1\%FS accuracy for near-atmospheric experiments. Finally, an oxygen gas sensor (Vernier) measures oxygen content from the exhaust before it is piped to the building ventilation system.

For any desired experimental chamber pressure, a certain mass flow of gas through the entire system is required to establish the desired flow velocity over the build area. The backpressure resulting from that flow in the chamber with valve 2 fully open must then be minimized to enable the largest possible operable range of pressures and flows in the system. In Figure \ref{fig:GasSystem}b, the system’s flow resistance is measured with the built-in flowmeter and pressure gauge and fit to a second-order curve. This curve is then extrapolated to show the full operating range of the system in Figure \ref{fig:GasSystem}c. Starting with valve 1 fully closed and the valve 2 fully open, opening valve 1 gradually moves the system's operating point outward along the blue curve, and closing valve 2 changes the resistance of the system, sending the blue curve upward to bring higher pressures into reach.

We use the van der Waals equation to characterize the relationship between pressure and mass flow required for adequate gas knife performance,
\begin{equation}
    \left[P+\frac{a}{\left(V/n\right)^2}\right]\cdot\left(V/n - b\right) = R\cdot T
    \label{eq:gas}
\end{equation}
Here, $P$ is the gas pressure (atm), $V$ is the gas volume (L), $n$ is the number of moles (mol), $R$ is the ideal gas constant (L-atm)/(mol-K), $T$ is the temperature (K), and $a$ and $b$ are empirical gas constants. Holding V constant, using $T$ = 298, $R$ = 0.08206 and empirical constants for argon ($a$ = 1.355 and $b$ = 0.03201), and solving iteratively for n, the mass flow of argon required to meet the steady-state velocity design objective at any given pressure can be calculated. This calculation is represented by the black solid line in Figure \ref{fig:GasSystem}c, originating at 0\,psig with 12\,NLPM based on nozzle performance.

\begin{figure}%[H]
    \centering
    \includegraphics[width=1.0\textwidth]{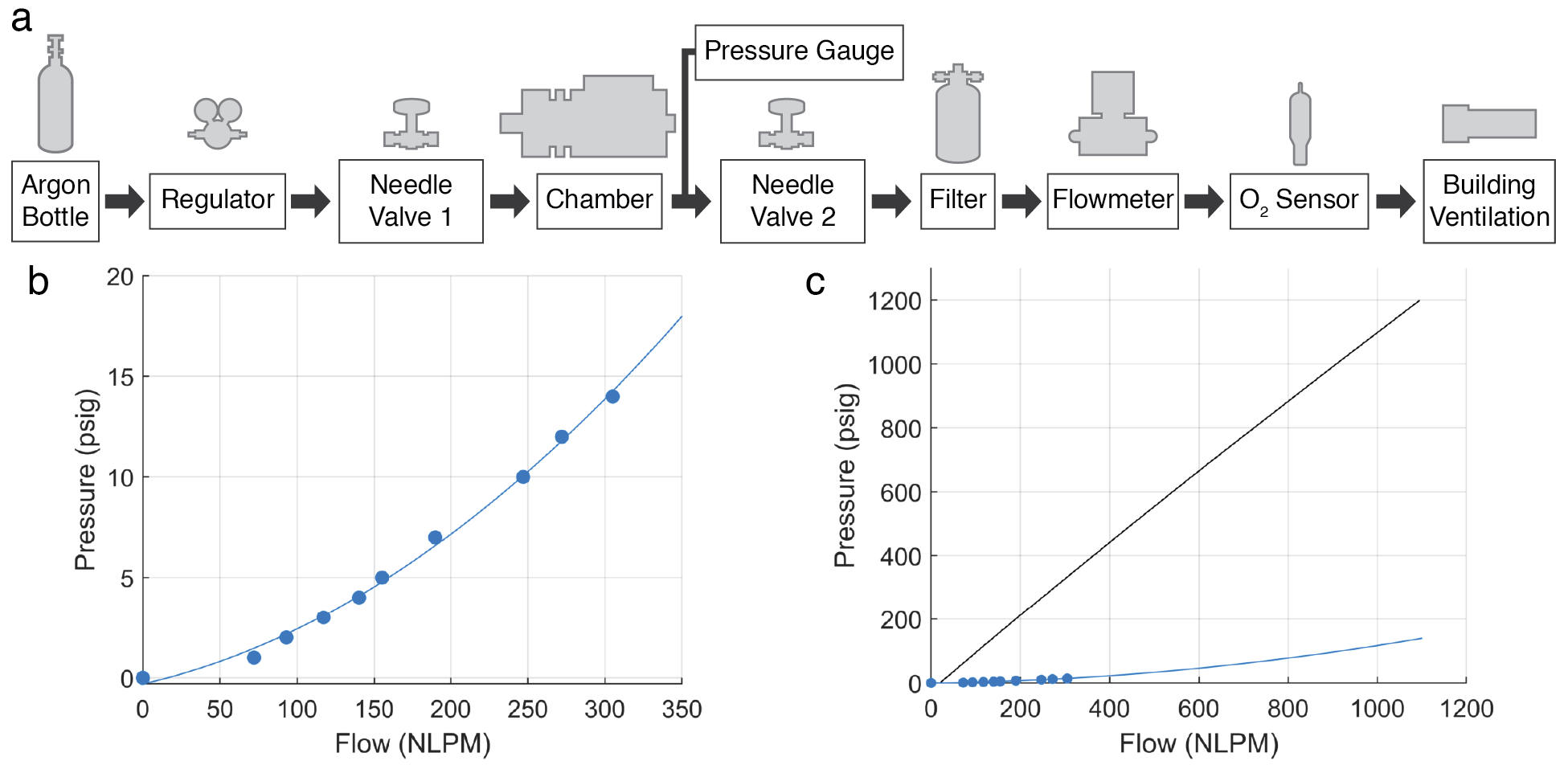}
    \caption{(a) Diagram of the gas delivery system. (b) Chamber pressure measurements taken at various mass flows of argon, fit with a second order equation. (c) Expected system flow resistance across the full operating range, using an extrapolated fit curve, with valve 2 fully open. The black line represents target experimental conditions, which can be reached by gradually closing valve 2 to raise the blue curve.}
    \label{fig:GasSystem}
\end{figure}

\subsection{Laser scanning system}

A custom-built mirror-based 2D scanning head is used to focus and direct the output of a fiber laser onto the build surface of the HLPM system. The laser (SP-0500-C-W-020-15-PIQ-011-001-001, SPI Lasers) produces 30\,-\,525\,W at a wavelength of 1075\,-\,1080$\mu$m. The laser’s output is directed via its output fiber into a collimator (106402X01, Coherent) where the beam is straightened. This beam then reflects off an angular alignment mirror (38-900, Edmunds Optics), passes through an AR-coated enclosure window (VPWW42-C, Thorlabs), and then reflects sequentially from a pair of orthogonal galvanometer mirrors (6240H, Novanta) before finally passing through an F-theta lens (FTH254-1064, Thorlabs), which focuses the straight beam down to a flat, horizontal focal plane below (see Figure \ref{fig:L-PBF_Def}a and \ref{fig:LaserSystem}a).

\begin{figure}%[H]
    \centering
    \includegraphics[width=1.0\textwidth]{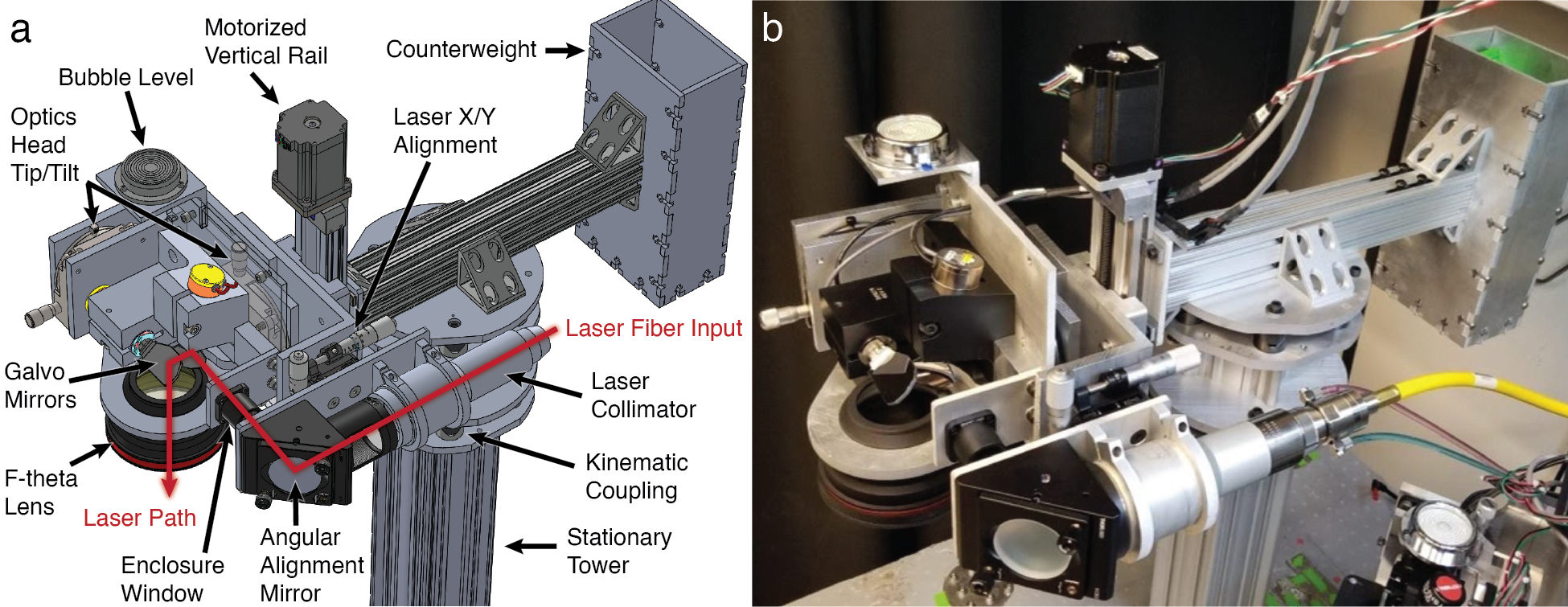}
    \caption{(a) CAD model and (b) image of the scanning apparatus.}
    \label{fig:LaserSystem}
\end{figure}

Unlike commercial L-PBF systems where the optics are fixed inside the upper portion of the machine to prevent dust exposure, a goal of the present design was to enable the scanning head to be precisely repositioned over multiple testbeds, including the HLPM apparatus. Therefore, the entire laser system is outfitted with kinematic couplings (VB-75-SM and CS-1125-CPM, Bal-tec) which allow for rotation in increments of 60\,degrees atop an aluminum extrusion tower (47065T501, McMaster Carr) attached to a rigid table. The laser scanning system and testbeds are enclosed behind a laser safety curtain. The galvanometer mirrors and F-theta lens comprising the scanning head are further enclosed (Figure \ref{fig:LaserSystem2}) in a custom stainless steel sheet metal box (Protocase). The enclosure is continuously purged with a 1\,LPM flow of high-purity nitrogen gas to prevent contamination of the optics inside and provide cooling to the galvanometer motors. The aforementioned enclosure window allows the laser to enter the enclosure with minimal reflective losses, while blocking the entrance of dust from the open lab environment. Three thermistors were installed behind the angular adjustment mirror and two galvanometer mirrors as a further safety measure, and the control program was set to deactivate automatically and provide notice should a thermistor be exposed to laser energy.

The galvanometers, F-theta lens, and enclosure (optics head) pivot jointly on two rotation stages (860-0150, Eksma Optics) whose axes of rotation intersect at the centroid of the first galvanometer mirror. This ensures that no matter the tip/tilt adjustments made by the user, the laser will strike the first galvanometer mirror at the center of its aperture. A top-mounted bubble level (2198A72, McMaster Carr) is used to level the optics head angle with an accuracy of approximately 1\,minute in the two rotation stage axes. The third axis of rotation is built rigidly into the overall construction, as it only affects the rotation angle of the projected laser focal plane, not its tip nor tilt, and can thus be easily corrected for in the software or in the positioning of the HLPM chamber.

The collimated laser beam must be precisely aligned to the galvanometer mirrors and F-theta lens to produce repeatable results. To this end, the laser fiber terminal, collimator, and angular alignment mirror are jointly suspended on a vertically oriented 2-axis stage (LX20, Thorlabs), which allows for precise translational alignment, and the angular alignment mirror is held in a kinematic mount (KCB2, Thorlabs) to give full control of the laser angle as it enters the galvanometer enclosure.

As the laser beam exits the F-theta lens, its diameter decreases toward a minimum focal diameter (beam waist) before expanding again. Viewed as a volume, the focusing laser energy would take the shape of a hyperboloid of one sheet, with the diameter changing linearly along the axis of laser travel except near the beam waist, where diffraction drives a non-linear relationship. The laser spot size can therefore be changed by translating the entire laser head assembly vertically using the system’s motorized vertical rail (SIMO, PBC Linear). The vertical rail’s carriage position is tracked using a Mitutoyo 500-171-30 digital caliper (McMaster Carr) with an accuracy of 0.025\,mm. One jaw of the caliper is bolted securely to the guide rail’s stationary frame and the other jaw is pulled down to touch the carriage for a highly repeatable measurement. Data correlating relative carriage position to laser spot size is presented in Figure \ref{fig:SpotSize}, where each data point is the average of five measurements. We compared our experimental data to an optical model created in OpticStudio 16.5 (Zemax) comprising the laser collimator and F-theta lens, and we noted that spot sizes assuming an ideal Gaussian-distributed laser input match our experimental values to within a 9\% difference in slope. The spot size measurement methodology is further explained in the Supporting Information. $z = 0$ corresponds to placement of the build surface at the beam waist, and negative laser head position corresponds to lowering the laser head toward the build platform. Although there appear to be two laser head positions which produce a given spot size, it is critical to operate only within $z < 0$ such that the beam waist is below the sample surface, not above. Operating with the beam waist (i.e. greatest energy density) above the build platform can rapidly heat or even ionize gas in the chamber, which could undesirably scatter or attenuate laser energy.

\begin{figure}[H]
    \centering
    \includegraphics[width=1.0\textwidth]{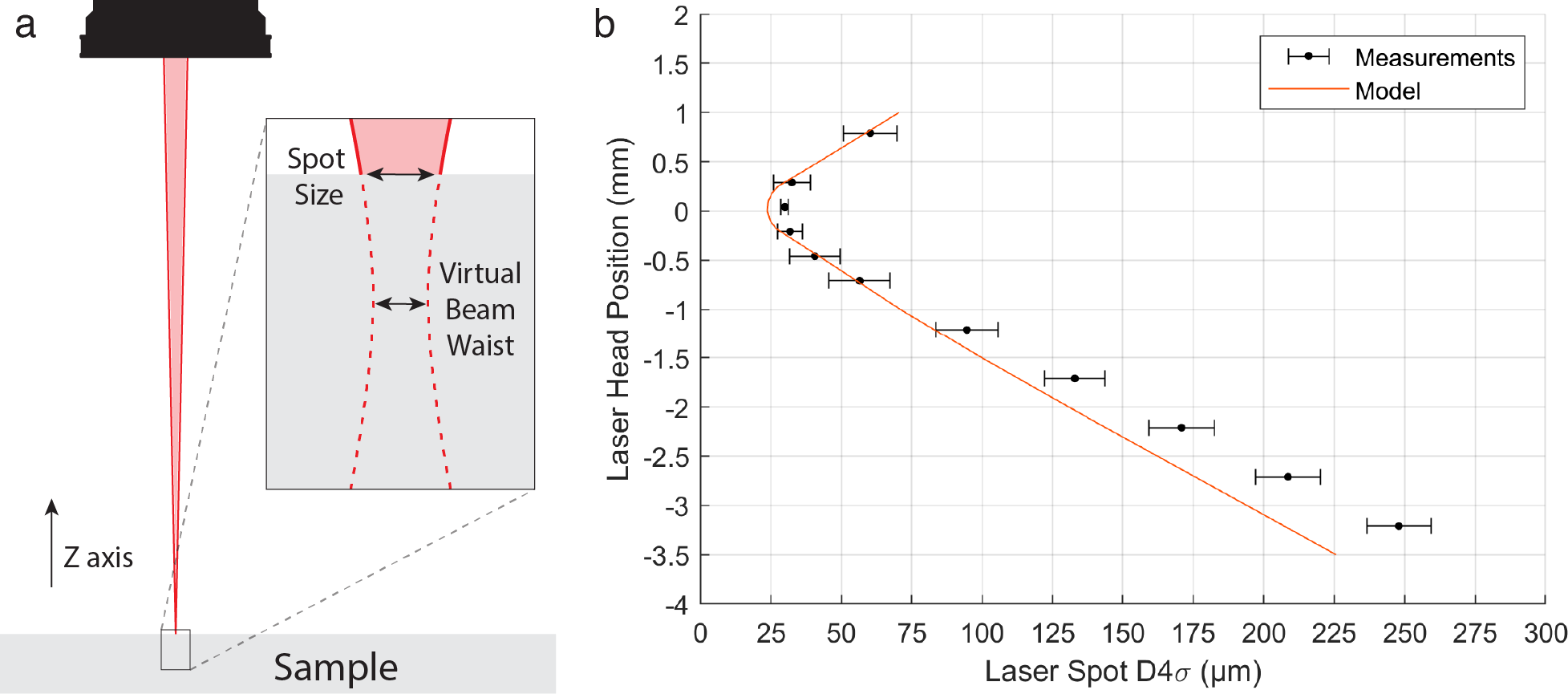}
    \caption{(a) Diagram (not to scale) of exemplary laser focus profile with the beam waist placed virtually below the sample surface. (b) Laser spot size versus relative vertical position $(z)$ of the scanning head.}
    \label{fig:SpotSize}
\end{figure}

The laser system is controlled using a cRIO 9039 FPGA (field programmable gate array) chassis from National Instruments, which communicates with the operator via ethernet to a dedicated PC workstation running LabVIEW software. A stepper driver module (NI-9503) actuates the vertical rail for laser focus adjustments, a digital IO module (NI-9375) communicates with the laser’s digital control hardware to operate it remotely, and an analog output module (NI-9269) sends voltages to synchronously control the two galvanometers and the laser power output at 10\,$\mu$s intervals. In order to conduct an experiment, a custom Python pre-processing script converts Gcode commands into a CSV file representing the required sequential voltage commands. When the system is activated, a custom LabVIEW program reads the CSV file and passes the values on to the FPGA, which in turn produces the synchronous galvanometer position and laser power signals. The PC workstation is also directly connected to the laser with a Serial USB adapter; this connection facilitates operation of the visible pilot laser beam, monitoring of the laser temperature, and manual command of the laser when necessary.

\section{System Validation and Experimental Results}

\subsection{Nominal melt track analysis}

As a first demonstration of HPLM system functionality, we show an array of vertical lines melted into a solid plate at 0\,psig. As noted on Fig.\,\ref{fig:SpotSize}, the D4$\sigma$ spot diameter at the plate surface is 109\,$\mu$m. In this first experiment, lines of constant speed (1\,m/s) and repeatedly alternating power (200\,W, 130\,W, 60\,W) were applied (perpendicular to the gas flow, moving upstream as the array is printed) with a center-to-center spacing of 1\,mm on a 0.25\,inch thick bare (i.e., no powder layer) SS316L plate. All substrate plates used in this study were prepared by sandblasting the top surface to enhance absorptivity, then cleaning with acetone. 

A high resolution height map of the melt track array was obtained using a laser confocal microscope (Keyence VK-X1050). The centerlines of all melt tracks were found to be within 10\,$\mu$m of their intended lateral location within the print area.

The position of the melt tracks is influenced by many factors, and thus any error in that spacing represents the accumulated contributions of errors associated with those factors. Important considerations include:
\begin{itemize}
    \item Alignment of all optical components (collimator, angular alignment mirror, galvos, F-theta lens, chamber optical window) must be maintained to ensure that no laser light is reflected or refracted away from the desired beam path across the full range of galvanometer rotation.
    \item Software must output properly scaled commands via the analog output module to produce appropriate galvanometer positioning.
    \item The analog output module must produce high fidelity analog signals to the galvanometer control board.
    \item The galvanometer control board must precisely convert voltage inputs into galvanometer positions.
    \item The F-theta lens must remain well aligned and clean to ensure a galvanometer's angular position translates precisely to the expected laser X,Y position.
    \item Parallelism of the build platform and laser focal plane is needed to ensure actual location of laser incidence on the plate is not distorted.
\end{itemize}

The ability to precisely direct the laser spot on the print bed is an important first step, but to delve deeper into system performance, one must inspect the melt track morphology. Nine sections were made in the plate as denoted by white lines in Figure \ref{fig:LaserVal}a. All sectioned samples were cut with a high speed diamond saw and mounted in copper or carbon powder, then sequentially ground and polished with 120, 300, 600, and 1200 grit sandpaper, a 3\,$\mu$m diamond slurry, and finally 0.05\,$\mu$m alumina suspension. The above procedure is according to the SumMet guide produced by Buehler \cite{Buehler2013BuehlerAnalysis}. All samples were etched with an aqua regia mixture: 1 part hydrochloric acid, 1 part nitric acid, and 1 part deionized water. The etched melt pool sections were imaged optically (Zeiss Smartzoom 5), and the resulting images were analyzed using a custom \textsc{Matlab} script to extract the depth, width, height, and cross sectional area. Further details on this script are given in the Supporting Information.

The resulting data points, shown in Figure \ref{fig:LaserVal}b, are the average profile measurements from three sections of each melt track, showing generally good uniformity across the array of melted lines. At 200\,W we see a slight quadratic variation in aspect ratio, from deeper, narrower melt pools near the center of the build area to shallower, wider pools at greater distance from the center. The variance in melt pool aspect ratio with respect to position on the plate increases with laser power, from $2.1\times10\textsuperscript{-3}$ at 60\,W to $4.4\times10\textsuperscript{-3}$ at 130\,W and $1.0\times10\textsuperscript{-2}$ at 200\,W. Similarly, the variance in melt pool area increases from $3.76\times10\textsuperscript{4}$\,$\mu$m$^2$ at 60\,W to $6.47\times10\textsuperscript{4}$\,$\mu$m$^2$ at 130\,W and $2.30\times10\textsuperscript{5}$\,$\mu$m$^2$ at 200\,W.

\begin{figure}[H]
    \includegraphics[width=1.0\textwidth]{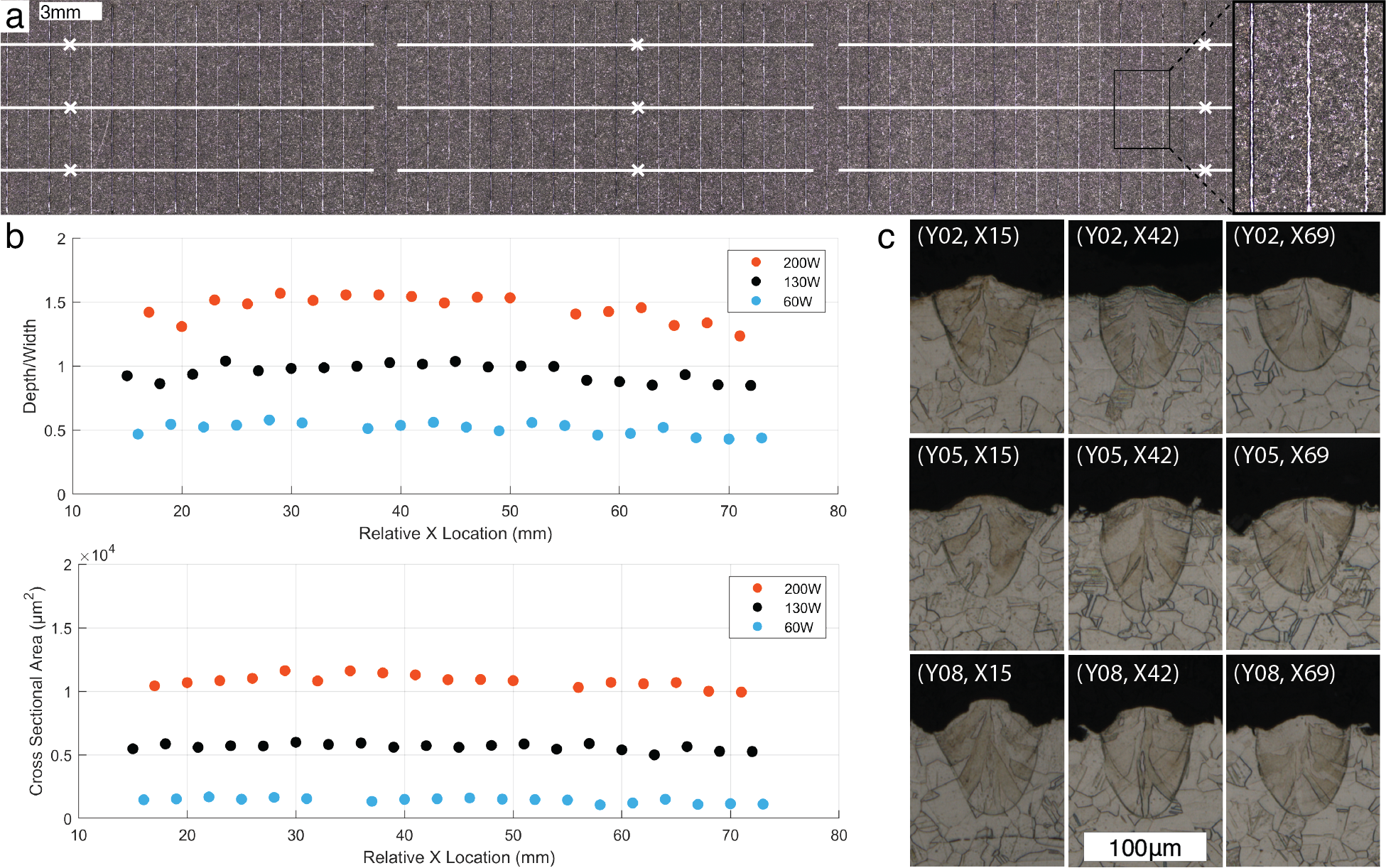}
    \caption{(a) Top down view of calibration melt tracks printed at 0\,psig with 1\,mm spacing. White horizontal lines denote section locations. A small area is magnified to the right for a clearer view of melt track surface quality. (b) Morphology data, taken as the average of three sections of each track. The gas knife nozzle terminates at X = 0 to the left of the print area shown, and the laser system origin (F-theta lens centerline) is located at X = 42. (c) Exemplary melt track section images from lines printed with a power of 130\,W. Coordinates are marked with corresponding white crosses. The Y position indicates section distance from the top of a track.}
    \label{fig:LaserVal}
\end{figure}

We attribute the trend at higher powers to intrinsic behavior of the F-theta lens, which causes the spot size to enlarge at positions farther away from the center of its aperture. As such, we measured the spot size when positioned at the laser system origin at X = 42\,mm to be 233.8\,$\mu$m and when at X = 3.9, 16.6, 29.3, 54.7, 67.4, 80.1, $\sigma$ = 251 241, 242, 234, 241, 244. Other effects such as variation in sample surface roughness, defects or contaminants in or on the sapphire pressure window and quartz slide, turbulence or aberrations in gas knife flow, and misalignment between the sample surface and laser focal plane may be present but were not considered to contribute meaningfully to the slight variation in melt pool aspect ratio and area. Therefore, we considered this acceptable performance to investigate melting behavior at elevated pressures.

\subsection{Laser melting at elevated pressure}

Next, experiments were performed at elevated pressure. Specifically, arrays of 8\,mm long melt tracks were printed on SS316L plate, with each track oriented perpendicular to the gas flow. A matrix of laser parameters ($P$ = 50, 150, 250, 350\,W, $\nu$ = 800, 1039, 1351, 1755, 2280, 2962, 3848, 5000\,mm/s) was sorted randomly and printed sequentially starting closest to the chamber outlet, and therefore incrementally farther upstream with respect to the flow direction. As such, the spatter and fumes from each melt track were not swept over the location of upcoming melt tracks. Six control lines ($P$ = 200\,W, $\nu$ = 4300\,mm/s) were added to the array at fixed positions across the entire print area in order to check for systematic errors in each experiment. Identical arrays were printed at 0\,psig, 150\,psig, and 300\,psig. The sample plates were then sectioned near their midpoints and characterized as described previously.

The resulting data are shown in Figure \ref{fig:BareTracks} in the form of melt track aspect ratio (depth/width), and cross-sectional area versus normalized enthalpy. Hann et al.\,\cite{Hann2011AParameters} first proposed normalized enthalpy as a means to study scaling of weld process parameters and bead (i.e, melt track) geometry. We use the following equation to define normalized enthalpy as
\begin{equation}
    \Delta H/h_s = \frac{AP}{\rho h_s\sqrt{\pi D\nu\sigma^3}},\ D = \frac{\kappa}{\rho c}
    \label{eq:normEnth}
\end{equation}
where $\Delta H$ is the change in enthalpy, $h_{s}$ is the specific enthalpy of melting, $A$ is laser absorptivity, $P$ is laser power, $D$ is thermal diffusivity, $\kappa$ is thermal conductivity, $\rho$ is density, $c$ is the specific heat, $\nu$ is the laser scan speed, and $\sigma$ is the laser spot diameter. We assume a constant value for absorptivity, $A = 0.4$. Values for enthalpy at melting, $h_{s} = 821\,kJ/kg$, density, $\rho = 7107\,kg/m^2$, and solidus temperature, $T_m = 1675\,K$, are used according to the measurements of Pichler et al.\,\cite{Pichler2020MeasurementsSteel}. We calculate $D = 7.34E-6\,m^2/s$ using additional values for thermal conductivity $\kappa = 35.56\,W/mK$ and specific heat $c = 681.9\,J/kgK$ at $T = 1675\,K$ reported by Kim \cite{Kim1975ThermophysicalSteels}.

\begin{figure}[H]
    \centering
    \includegraphics[width=1.0\textwidth]{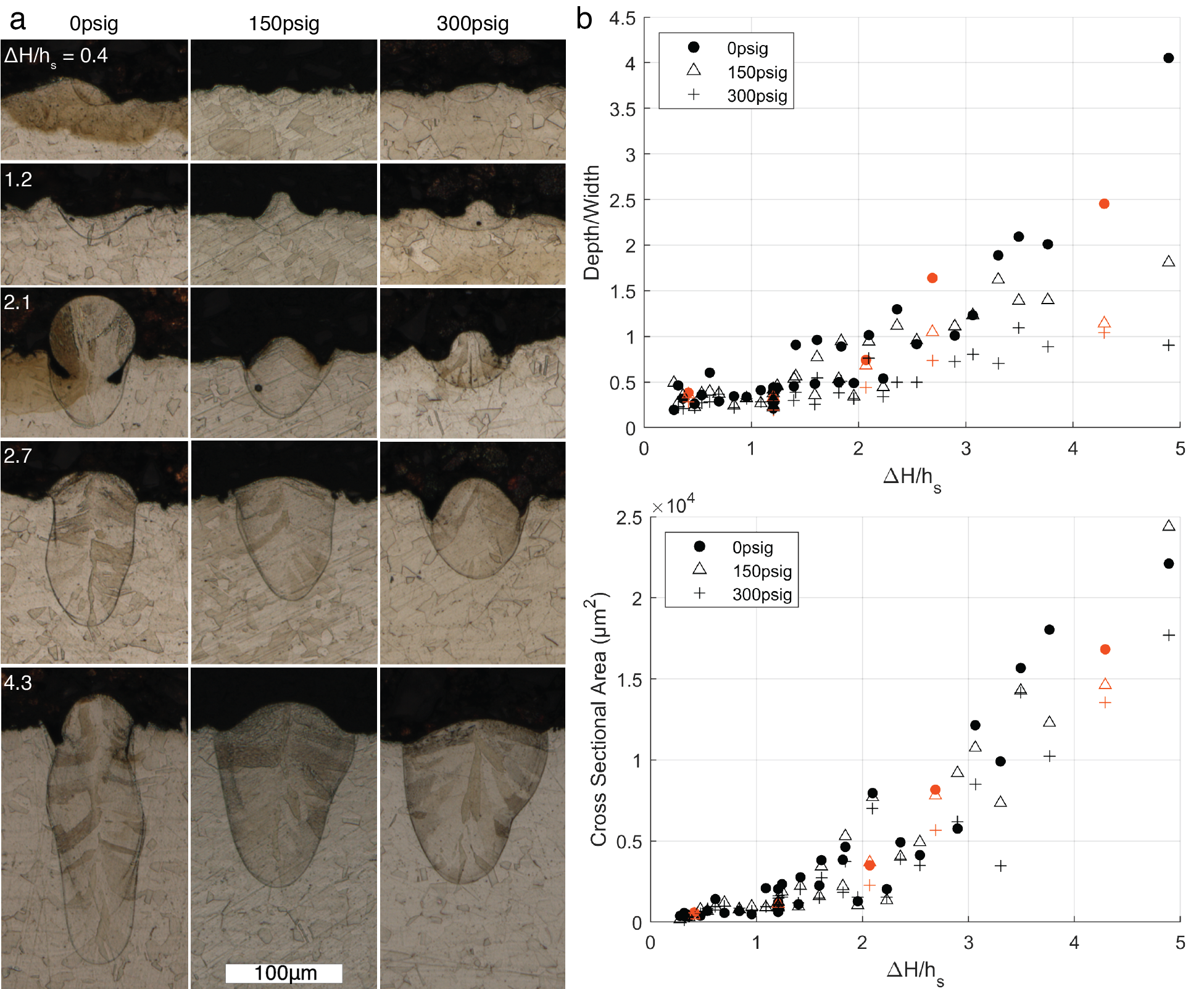}
    \caption{(a) Optical cross section views of individual exemplary melt tracks printed in SS316L plate. Normalized enthalpy values are listed for each row of images, and pressure values for each column. Images are at equal scale. (b) Melt pool depth/width ratio and total cross sectional area of each individual melt track, with exemplary tracks shown in (a) marked as red data points.}
    \label{fig:BareTracks}
\end{figure}

As expected, we see in Figure \ref{fig:BareTracks} a trend of increasing aspect ratio with increasing normalized enthalpy. Traditionally in laser welding, $d/w = 0.5$, i.e., when the depth exceeds the half-width of the melt pool, is considered the threshold between conduction and keyhole mode melting \cite{Eagar1983TemperatureSources.,King2014ObservationManufacturing}. Formation of keyhole porosity is a complex process which occurs at some point beyond the aspect ratio threshold, and parabolic melt pool shapes are considered desirable in L-PBF towards remelting into the previous layer, as long as keyhole pores are not formed.

At elevated pressure, melt pool penetration depth decreases and width increases, relative to the nominal results at 0\,psig. This is consistent with sub-atmospheric studies of L-PBF and laser welding experiments, where weld depth increased and width decreased with lower pressure \cite{Katayama2011DevelopmentVacuum, Jiang2017EffectWelding,Li2018ExperimentalPressure,Jiang2019ComparisonSub-atmosphere,Sokolov2015ReducedSteel,Bidare2018LaserPressures}. In the present elevated pressure experiments, the normalized enthalpy at which $d/w$ exceeds 0.5 grows slightly with pressure (e.g., approximately $\Delta H/h_{s}$ = 1.6 at 150\,psig, and $\Delta H/h_{s}$ = 2.1 at 300\,psig), and the deviation in $d/w$ grows with increased normalized enthalpy. Yet, no significant shift in melt track sectional area is observed. The preliminary results here suggest that ambient pressure manipulates the force balance at the melt pool surface, which in turn influences the melt track geometry and progression of aspect ratio with increasing applied energy. However, under these pressures the amount of energy transferred to the melt pool is relatively unchanged, supporting our conclusion that the gas flow within the HPLM system is sufficient to prevent laser attenuation due to plume dynamics. Investigation of keyhole pore formation statistics versus pressure is left as future work.

As a final demonstration of the HPLM system in the specific context of L-PBF, an array of melt tracks was printed on a 76\,$\mu$m thick powder bed at both 0\,psig and 150\,psig. Gas atomized SS316L powder (John Galt Steel) was used, with a particle size distribution of 15-45\,$\mu$m reported by the supplier. The underlying SS316L sample plate was prepared as before for bare plate studies, then bolted to the build platform with the addition of a thin steel shim which sets the powder layer thickness by forming a well in the print area. The shim was cut to fit the sample plate bolt pattern and desired powder deposition area, then lapped to remove burrs and reach a uniform thickness of 76\,$\pm$\,3\,$\mu$m. Once the sample plate and shim were bolted to the platform, powder was dispensed manually to the mid-line of the sample well, and a 0.125\,inch thick machinist's edge was held against the shim and drawn length-wise across the print area to spread the powder. A flashlight was shone at a slight angle across the powder bed in two orthogonal directions to check for unacceptable streaks or defects in the powder surface. An illustration of this procedure is given in the Supporting Information. At 0\,psig the gas knife performed as designed, reaching the desired flow rate without disturbing powder. At 150\,psig a 40\% reduction in mass flow was required to prevent disturbance of the powder bed during processing. Resulting powder bed melt track properties at 0\,psig and 150\,psig are shown in Figure \ref{fig:PowderTracks}.

\begin{figure}[H]
    \centering
    \includegraphics[width=1.0\textwidth]{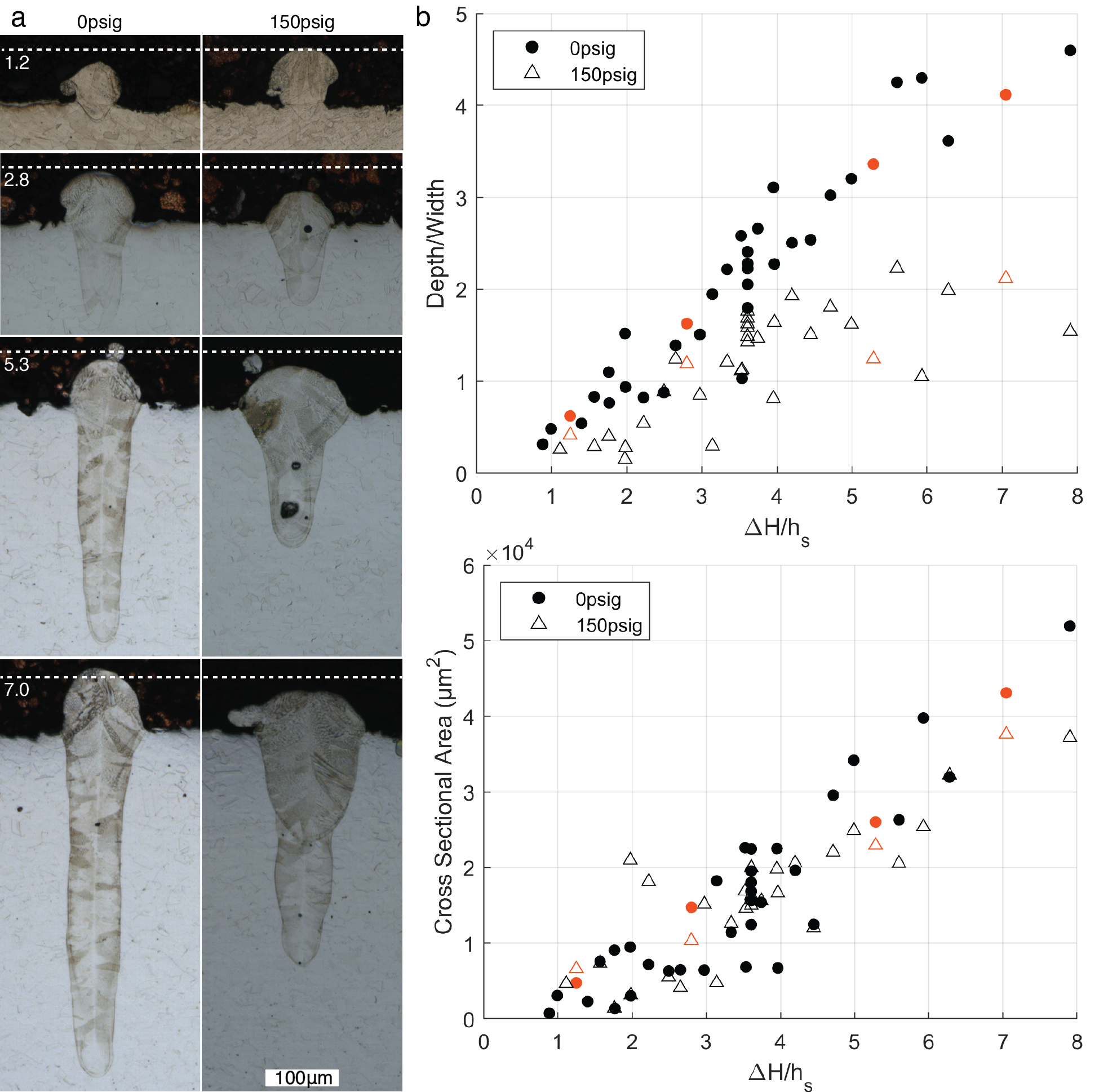}
    \caption{(a) Optical section views of individual exemplary melt tracks melted into a 76\,$\mu$m thick layer of SS316L powder. Normalized enthalpy values are listed for each row of images, and pressure values for each column. Images are at equal scale, and the powder layer thickness is denoted by dotted white lines. (b) Melt pool depth/width ratio and total total cross sectional area of each individual melt track, with exemplary tracks marked in red.}
    \label{fig:PowderTracks}
\end{figure}

Just as observed with bare plate experiments, increasing pressure produced shallower, wider melt tracks. Melt track sectional area remains consistent with increasing pressure, while the melt pool aspect ratio significantly changes at 150\,psig. Melt tracks transitioned from conduction mode to keyhole mode at similar values of $\Delta H/h_{s}$ as with bare plate; this was approximately $\Delta H/h_{s}$ = 1.1 at 0\,psig and $\Delta H/h_{s}$ = 2.0 at 150\,psig. To investigate the negative impact of reduced gas flow on the resulting line array, the same reduction in gas flow was repeated for melting experiments on a SS316L bare plate and no change in melt track dimensions was observed relative to the nominal flow bare plate experiments. Taken together, our findings suggest that pressure has a significant influence on melt pool geometry in L-PBF, and that the HPLM system is capable of precision investigation of L-PBF at pressures of up to 150\,psig, considering both machine operation and gas flow dynamics.

To achieve high pressure multi-layer L-PBF, a larger chamber is required to allow room for a vertical stage and recoating mechanism without impinging on the clear path required for a stable, effective gas knife. Critically, greater distance between the powder bed and the laser optical window above will reduce the danger of direct impacts of spatter on the window and greatly alleviate the consequences of recirculation. Finally, a gas system which reuses inert gas via a continuous circuit will be required, as at the highest pressures currently reachable by the HPLM system (1200\,psig), a standard bottle of argon could be drained in as little as 10\,minutes, producing a very short window within which to conduct experiments.

\section{Conclusion}
This paper presents the design and validation of an instrument that enables research on the influence of elevated ambient pressure on laser materials processing, specifically guided to laser powder bed fusion and laser welding. Exemplary bare plate and single layer powder bed experiments demonstrate the testbed’s performance at pressures up to 300\,psig. The HPLM system will enable direct observation of the impact of elevated pressure on the process windows of commonly processed materials, and will enable the exploration of new materials which cannot currently be processed with sufficient quality in L-PBF.

\section{Author Contributions: CRediT taxonomy}
\textbf{David A.\,Griggs:} Conceptualization, Methodology, Software, Validation, Formal analysis, Investigation, Resources, Data curation, Writing – original draft, Writing – review and editing, Visualization, Project administration. 
\textbf{Jonathan S.\,Gibbs:} Conceptualization, Software, Resources, Writing - original draft. 
\textbf{Stuart P.\,Baker:} Conceptualization, Software, Resources. 
\textbf{Ryan W.\,Penny:} Conceptualization, Validation, Formal analysis, Writing - review and editing. \textbf{Martin C.\,Feldmann:} Conceptualization. 
\textbf{A.\,John Hart:} Conceptualization, Methodology, Writing – review and editing, Visualization, Supervision, Project administration, Funding acquisition.

\section{Acknowledgements}
Financial support for this project was provided by Honeywell Federal Manufacturing \& Technologies (Honeywell FM\&T). We thank Paul Carson (MIT), Dan Gilbert (MIT), Joe Wight (MIT), Ben Brown (Honeywell FM\&T), and Rachel Grodsky (Honeywell FM\&T) for their valuable input. J.S.G.~and S.P.B.~also acknowledge financial support of their graduate studies from the United States Navy and MIT Lincoln Laboratory, respectively. 

\appendix
\counterwithin{figure}{section}
\section{Supporting Information}
\subsection{Laser power measurement}
There will necessarily be some amount of laser power lost to reflections or scattering in the optical components. The laser's power output was measured with all relevant optics in the beam path (including the sapphire pressure window and quartz slide), such that all optical losses that would occur in normal use are directly represented. Measurements at laser power commands from 30\,W to 150\,W were made with an Ophir Starbright meter and 30(150)A-LP1-18 sensor. The high linearity of these measurements $(R^{2} = 1)$ allows for extrapolation of expected laser power at the sample plate across the entire power range of the laser, from 30\,W to 500\,W. Thus, all mentions of laser power in this paper refer to the actual laser power which reaches the sample.

\begin{figure}[H]
    \centering
    \includegraphics[width=0.5\textwidth]{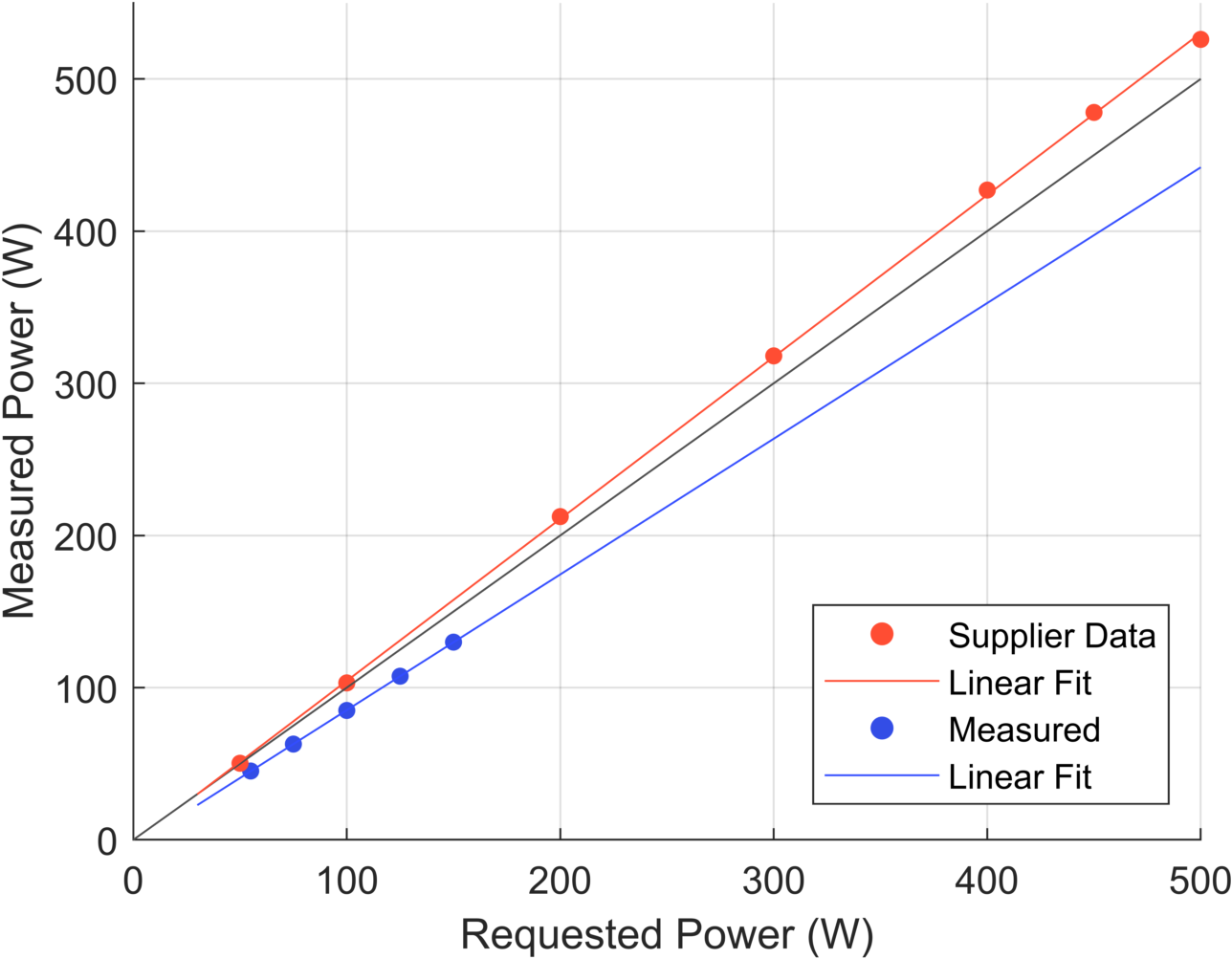}
    \caption{Requested laser power vs. actual laser power before (Supplier Data) and after (Measured) optical losses.}
    \label{fig:LaserPower}
\end{figure}

\subsection{Laser spot size measurement}
The spot size of the laser was imaged at various laser head relative positions using a DMM 37UX226-ML CMOS camera (The Imaging Source) with a pixel size of 1.85$\mu$m$\times$ 1.85$\mu$m, and IC Capture software (The Imaging Source). The sensor is mounted directly to the HPLM build platform at a height measured precisely with a dial indicator, such that measurements can be translated directly to any sample of known thickness. The laser is fired in low-power mode at a minimum output of 30\,W. The angular alignment mirror is replaced with a beam sampler (BSF20-C, Thorlabs) such that approximately 90-99\% of the energy (depending on polarization) is transmitted to a beam dump, while the remaining laser energy enters the galvanometer enclosure and is focused by the F-theta lens as in normal operation. A combination of reflective and absorptive neutral-density filters (Thorlabs) are placed in the focusing beam path to reduce the energy further to a brightness appropriate for the CMOS sensor. Images are taken at a rate of twenty frames per second both to capture dark frames before the laser engages and the spot appears, but also in order to capture the laser spot quickly and thus minimize the impact of thermal warping of the ND filters on spot measurements. A \textsc{Matlab} script was created to subtract dark energy from resulting images, then locate the laser spot centroid and plot a two-dimensional Gaussian fit to determine the D4$\sigma$ diameter of the laser beam. The laser used in the HPLM has a Gaussian energy distribution with an M$^{2}$ value of 1.1\,$\pm$\,0.1. The resulting D4$\sigma$ measurements, shown in Figure \ref{fig:SpotSize}c, allow for repeatable spot size selection at the surface of any desired sample of known thickness.

\subsection{Laser auxiliary systems}
A thermal safety system was devised in case the angular alignment mirror or galvanometer mirrors should fail while the laser is in use. Three thermistors (BC2385-ND, Digikey) are affixed behind the mirrors atop sandblasted aluminum strike-plates, such that laser energy penetrating the mirror would quickly warm the thermistor and/or surrounding aluminum. The thermistors are wired in parallel voltage divider circuits to the FPGA via an analog input module (NI-9205, National Instruments). The FPGA continuously monitors each voltage input and will quickly shutdown the system should any voltage signal change beyond a set threshold. The completed laser tower is presented in Figure \ref{fig:LaserSystem2} below.
\begin{figure}[H]
    \centering
    \includegraphics[width=0.7\textwidth]{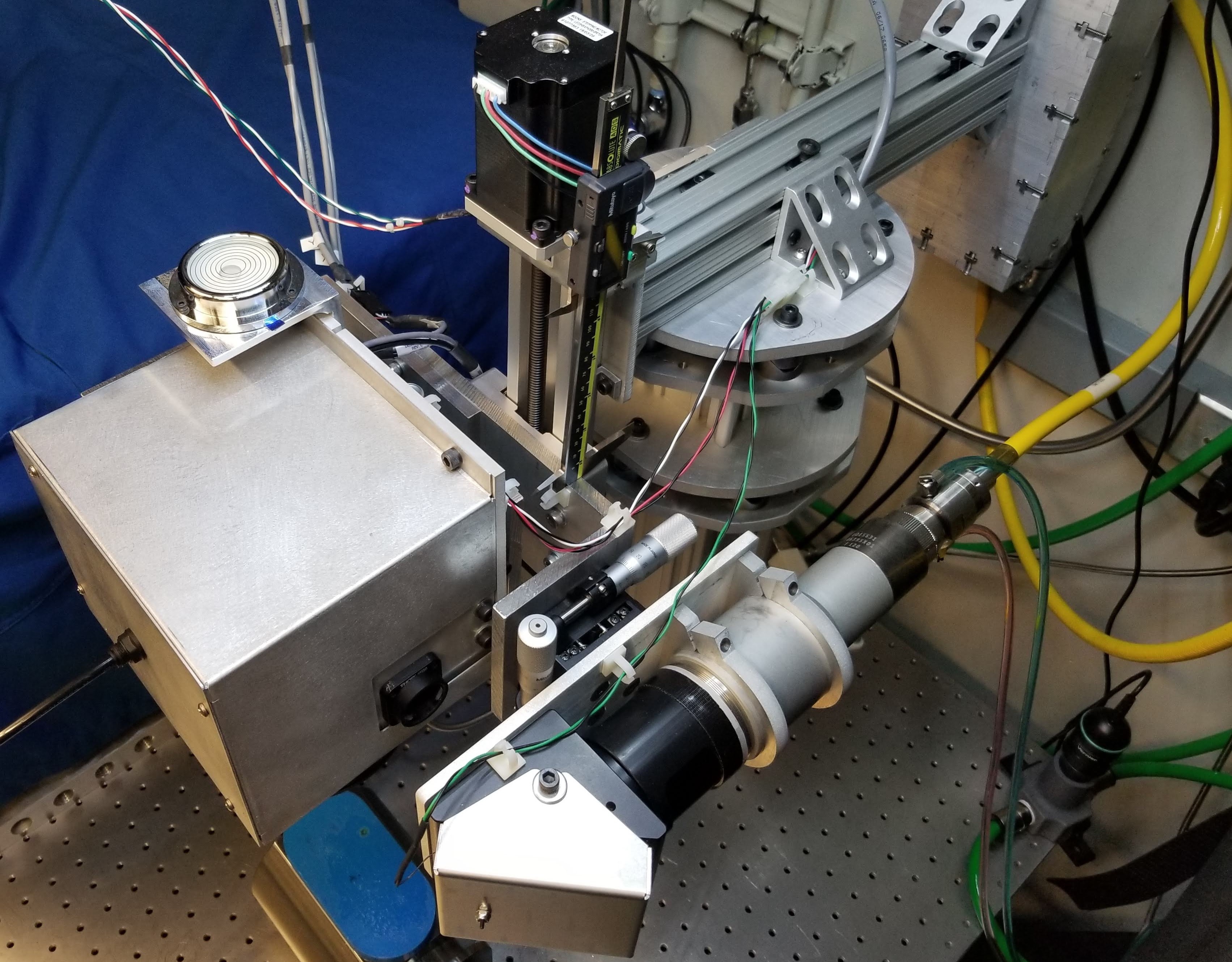}
    \caption{Image of the complete laser system with enclosure installed, and with nitrogen gas supply tubing and thermistor wiring (white/black, red/black, green/black) in place.}
    \label{fig:LaserSystem2}
\end{figure}

\subsection{Powder bed sample preparation and monitoring}
The detailed process of preparing a single powder layer is portrayed in Figure \ref{fig:powderPrep} below. The shim and sample plate are bolted to the build platform, then powder is spread in a line along the centerline of the long axis of the sample plate, and finally the powder is spread with a machinists edge in one quick, uniform motion.

\begin{figure}[H]
    \centering
    \includegraphics[width=0.8\textwidth]{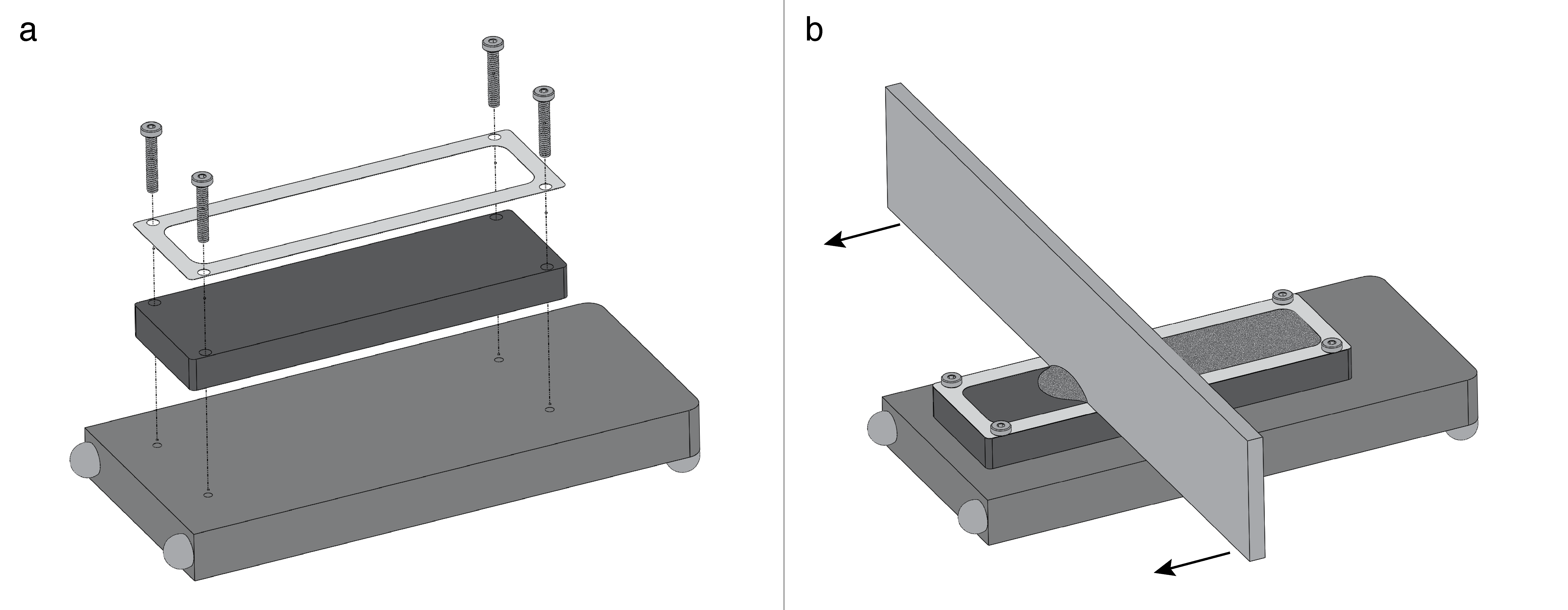}
    \caption{Powder bed sample preparation process. (a) Exploded view of sample plate assembly. (b) schematic of powder spreading action.}
    \label{fig:powderPrep}
\end{figure}

Figure \ref{fig:loading} demonstrates the process of loading a sample into the pressure chamber.

\begin{figure}[H]
    \centering
    \includegraphics[width=0.8\textwidth]{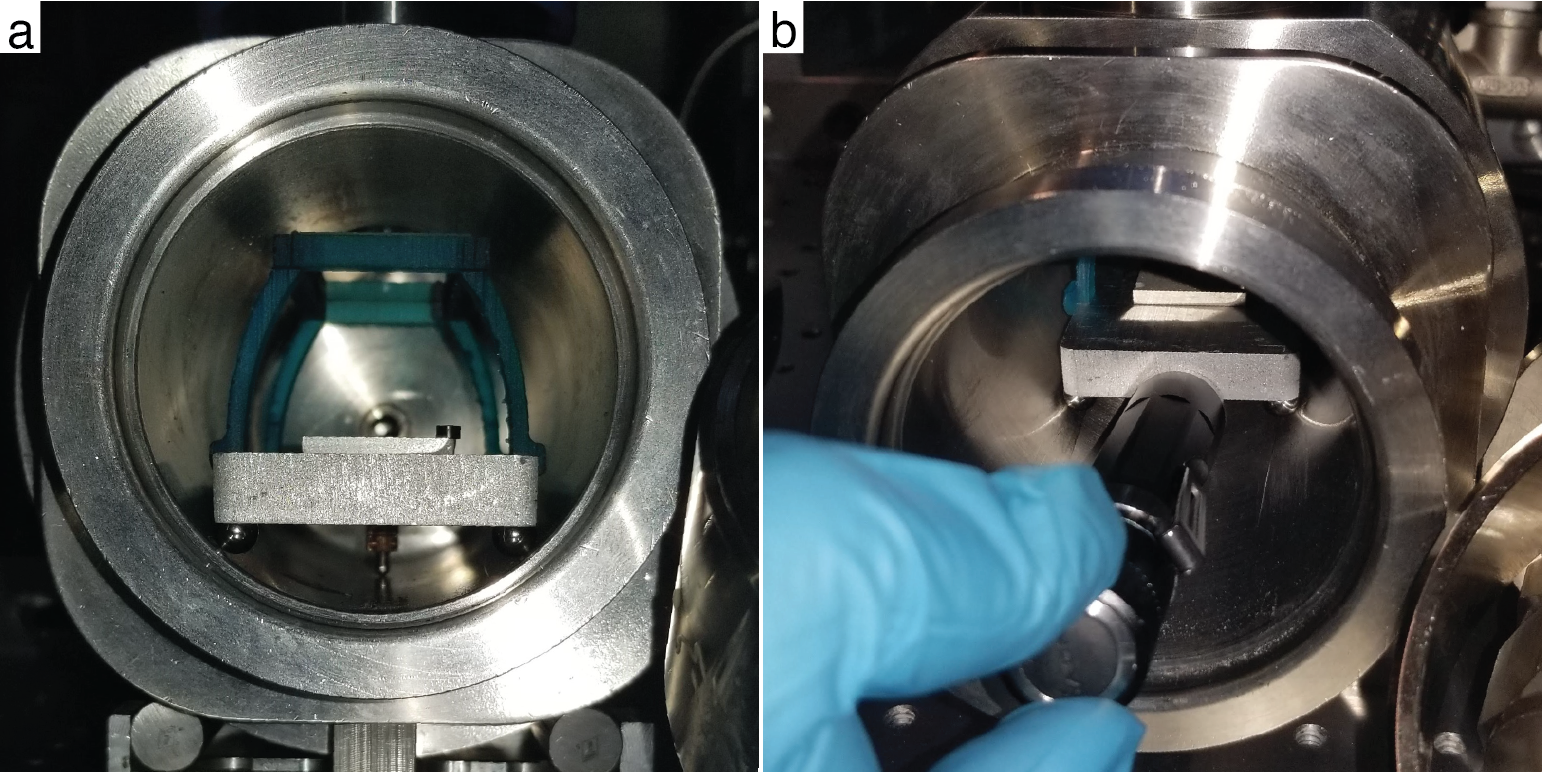}
    \caption{(a) Initial placement of build platform. (b) Final alignment of build platform, using a rigid object with a single point of contact to minimize lateral forces.}
    \label{fig:loading}
\end{figure}

During processing, the powder bed is continuously monitored via a machine vision camera by the operator, such that any scattering of powder can be immediately noted and the experiment terminated. Exemplary operator views are shown in Figure \ref{fig:camView}.

\begin{figure}[H]
    \centering
    \includegraphics[width=0.6\textwidth]{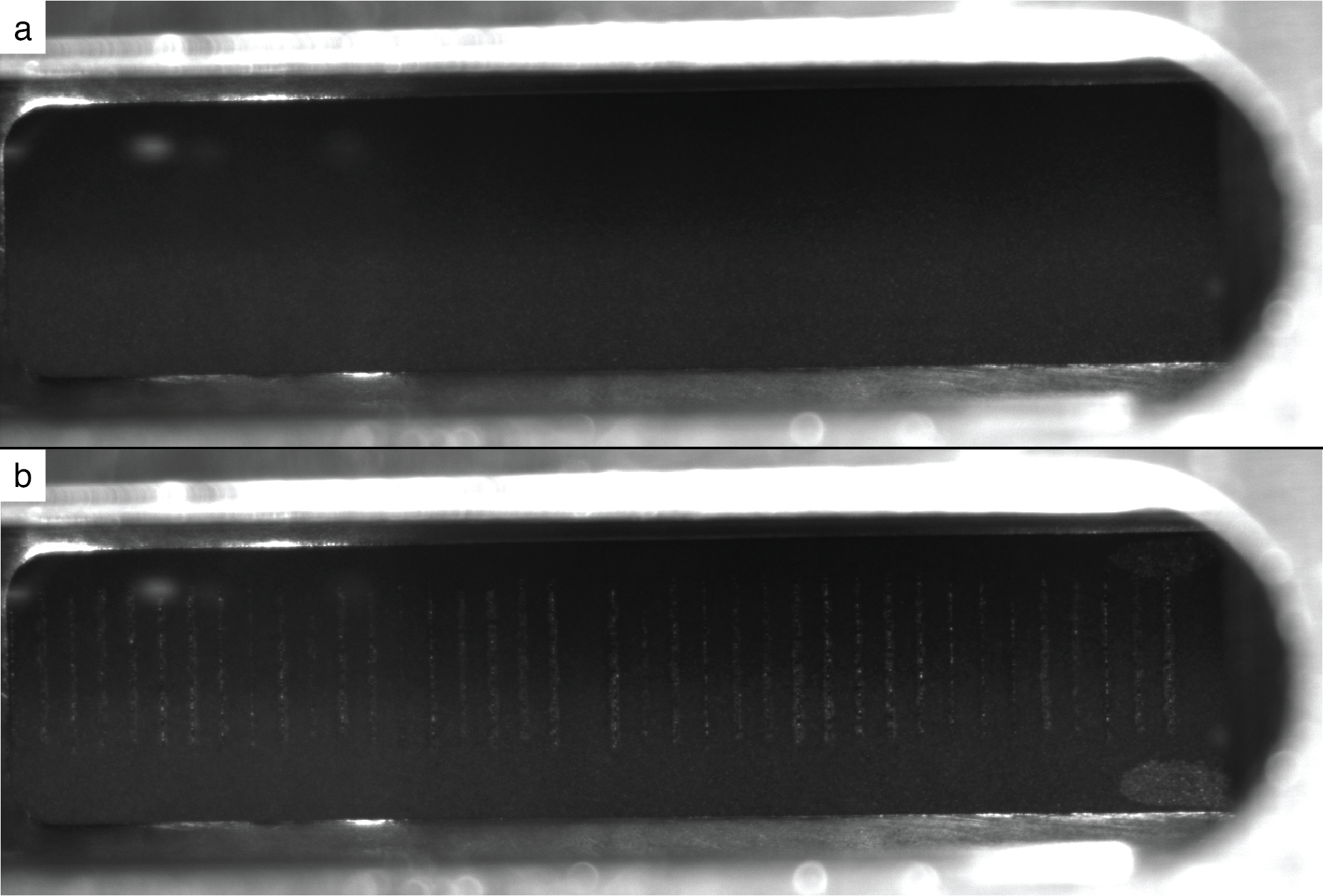}
    \caption{Exemplary images of powder bed as monitored by the machine vision camera (a) before and (b) after a print.}
    \label{fig:camView}
\end{figure}

\subsection{Comparison of experimental data}
\begin{figure}[H]
    \centering
    \includegraphics[width=1.0\textwidth]{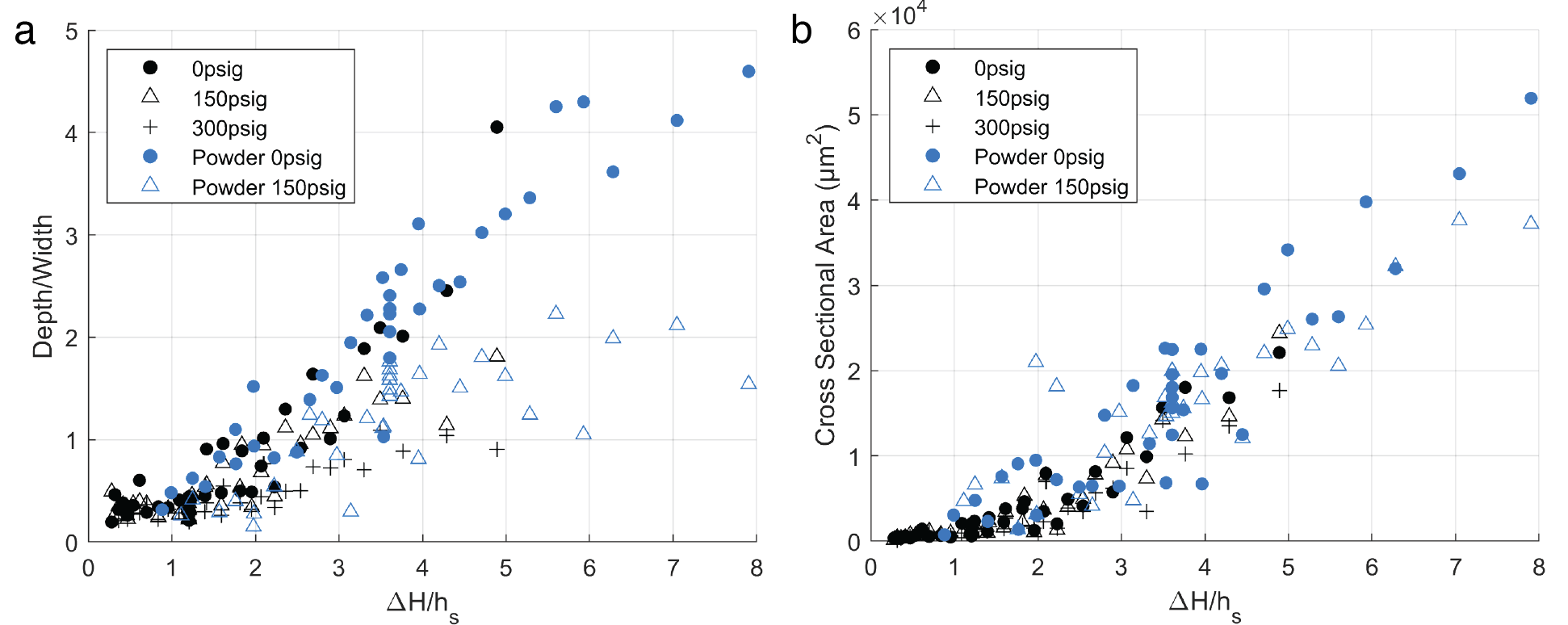}
    \caption{(a) Depth to width ratio and (b) cross sectional area data from the bare plate and powder layer experiments discussed in Results are overlaid.}
    \label{fig:bareVsPowder}
\end{figure}

\subsection{Melt track analysis}
To standardize the analysis of melt track geometry, a semi-automated routine was developed to identify the boundaries of the melt track and measure important track parameters. To efficiently accommodate a wide range of melt-track aspect ratios with a minimal number of measurements, we employ numerical integration using Simpson's first method in polar coordinates. This method lends itself well to dividing the surface and sub-surface track geometry which frequently diverge in size and shape. The program employs a masking technique to remove the upper dark region (the mounting media) in the section image and the lower light region (the undisturbed plate) below the melt-track. Further, the surface level was estimated and a rotational transformation was applied to the image to correct for any rotational misalignment of the base plate in the section image. Figure \ref{fig:SectionAnalysis} depicts this level surface estimate as a red line.  The melt track is then identified by the analyst using a separate region-of-interest bounding box for ``above grade'' and the ``below grade'' regions. An array of 32 guide lines, evenly spaced $\Delta\theta = 11.25$\textdegree\ apart, are superimposed on the image radiating outward from the centers of these two boxes where they meet the grade line to aide in identifying each radii to the top or bottom boundary of the melt track. 

\begin{figure}[H]
    \centering
    \includegraphics[width=1.0\textwidth]{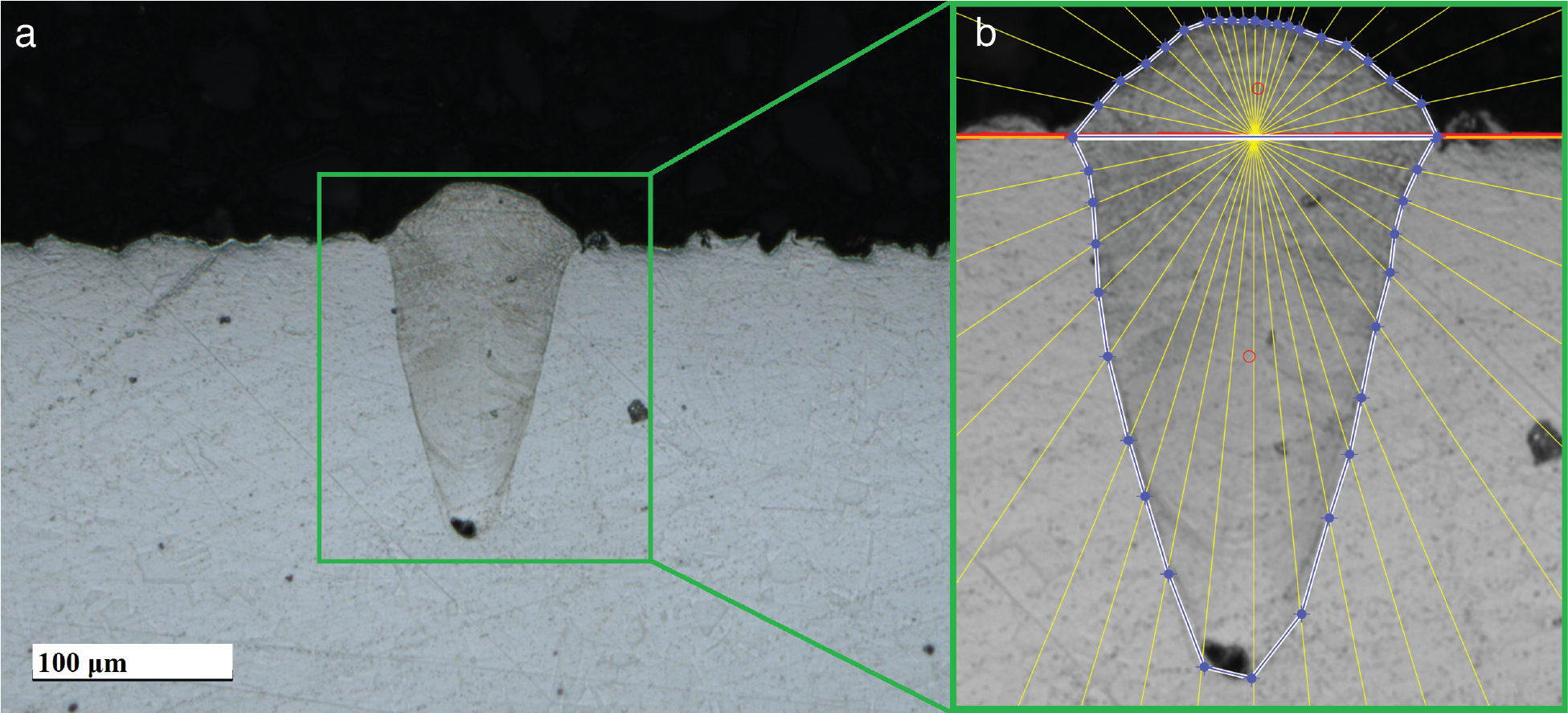}
    \caption{Example of graphical melt track section analysis. (a) Optical micrograph of 316L bare plate scanned with laser power of 400\,W, scan speed of 1.27\,m/s ($\Delta H/h_{s}$ = 36.9) at pressure of 150\,psig. (b) The same track image processed with above grade and below grade regions outlined in blue.}
    \label{fig:SectionAnalysis}
\end{figure}

Radii are first estimated using the light-dark transitions present in the masked image and then manually adjusted by eye by the analyst.  The parameters obtained include depth of the melt track below the plate grade, height of the track above the grade, width at grade, width of the above grade (more used in powder-bed studies) and the areas of each region, $A$. $A$ is approximated numerically by applying Simpson's first method in polar coordinates using the $n = 17$ radii, $\rho_i$, obtained for each angle of the shape as follows:

\begin{equation}
    A = \frac{1}{2}\int_0^{\pi/2} \rho(\theta)^2\:d\theta \approx\frac{1}{2}\sum_{i=1}^n \frac{\Delta\theta}{3}\left(\rho_i\right)^2 \cdot S.M._i%}_{\textrm{Simpson's 1$^{\textrm{st}}$ Method}}
    \label{eq:SimpsonsArea}
\end{equation}

where the Simpson's Multipliers, $S.M._i$, are $\left\{1, 4, 2, 4, 2, \cdots, 2, 4, 2, 4, 1\right\}$. Additionally, centroid coordinates $(\overline{\rho},\overline{\theta})$ (displayed in Figure \ref{fig:SectionAnalysis} as red circles) are calculated as follows:

\begin{gather}
    \overline{\rho} \approx \frac{\sum_{i=1}^n \left(\rho_i\right)^3 \cdot S.M._i}{\sum_{i=1}^n \left(\rho_i\right)^2 \cdot S.M._i}\\
    \overline{\theta} \approx \frac{\sum_{i=1}^n \theta_i\cdot\left(\rho_i\right)^2 \cdot S.M._i}{\sum_{i=1}^n \left(\rho_i\right)^2 \cdot S.M._i}
    \label{eq:SimpsonsCentroids}
\end{gather}

For high aspect ratios, such as instances of severe keyhole morphology, this method was adapted slightly to include 4 additional radii at half the global angular step-size. This may be seen in Figure \ref{fig:SectionAnalysis} in the increased density of points (and guide lines) at the bottom and top of the melt track.

\end{flushleft}
\bibliography{main}

\begin{thebibliography}{10}

\bibitem{Meier2017ThermophysicalExperimentation}
Christoph Meier, Ryan~W. Penny, Yu~Zou, Jonathan~S. Gibbs, and A.~John Hart.
\newblock {Thermophysical phenomena in metal additive manufacturing by
  selective laser melting: Fundamentals, modeling, simulation and
  experimentation}.
\newblock {\em Annual Review of Heat Transfer}, 20:241--316, 2017.

\bibitem{Solana1997AMaterials}
Pablo Solana and Guillermo Negro.
\newblock {A study of the effect of multiple reflections on the shape of the
  keyhole in the laser processing of materials}.
\newblock {\em Journal of Physics D: Applied Physics}, 30(23):3216--3222, 1997.

\bibitem{Gusarov2009Two-dimensionalMelting}
A.~V. Gusarov and I.~Smurov.
\newblock {Two-dimensional numerical modelling of radiation transfer in powder
  beds at selective laser melting}.
\newblock {\em Applied Surface Science}, 255(10):5595--5599, 2009.

\bibitem{Boley2017CalculationManufacturing}
C.~D. Boley, Saad~A. Khairallah, and Alexander~M. Rubenchik.
\newblock {Calculation of laser absorption by metal powders in additive
  manufacturing}.
\newblock {\em Additive Manufacturing Handbook: Product Development for the
  Defense Industry}, 54(9):507--517, 2017.

\bibitem{Trapp2017InManufacturing}
Johannes Trapp, Alexander~M. Rubenchik, Gabe Guss, and Manyalibo~J. Matthews.
\newblock {In situ absorptivity measurements of metallic powders during laser
  powder-bed fusion additive manufacturing}.
\newblock {\em Applied Materials Today}, 9:341--349, 2017.

\bibitem{Khairallah2020ControllingPrinting}
Saad~A. Khairallah, Aiden~A. Martin, Jonathan~R.I. Lee, Gabe Guss, Nicholas~P.
  Calta, Joshua~A. Hammons, Michael~H. Nielsen, Kevin Chaput, Edwin Schwalbach,
  Megna~N. Shah, Michael~G. Chapman, Trevor~M. Willey, Alexander~M. Rubenchik,
  Andrew~T. Anderson, Y.~Morris~Wang, Manyalibo~J. Matthews, and Wayne~E. King.
\newblock {Controlling interdependent meso-nanosecond dynamics and defect
  generation in metal 3D printing}.
\newblock {\em Science}, 368(6491):660--665, 2020.

\bibitem{Khairallah2016LaserZones}
Saad~A. Khairallah, Andrew~T. Anderson, Alexander Rubenchik, and Wayne~E. King.
\newblock {Laser powder-bed fusion additive manufacturing: Physics of complex
  melt flow and formation mechanisms of pores, spatter, and denudation zones}.
\newblock {\em Acta Materialia}, 108:36--45, 4 2016.

\bibitem{Sun2017PowderOverview}
S.~Sun, Milan Brandt, and M.~Easton.
\newblock {\em {Powder bed fusion processes: An overview}}.
\newblock Elsevier Ltd, 1 2017.

\bibitem{Pei2017NumericalPowder}
Wei Pei, Wei Zhengying, Chen Zhen, Li~Junfeng, Zhang Shuzhe, and Du~Jun.
\newblock {Numerical simulation and parametric analysis of selective laser
  melting process of AlSi10Mg powder}.
\newblock {\em Applied Physics A: Materials Science and Processing},
  123(8):1--15, 2017.

\bibitem{ScipioniBertoli2017OnMelting}
Umberto Scipioni~Bertoli, Alexander~J. Wolfer, Manyalibo~J. Matthews, Jean
  Pierre~R. Delplanque, and Julie~M. Schoenung.
\newblock {On the limitations of Volumetric Energy Density as a design
  parameter for Selective Laser Melting}.
\newblock {\em Materials and Design}, 113:331--340, 2017.

\bibitem{Matthews2016DenudationProcesses}
Manyalibo~J. Matthews, Gabe Guss, Saad~A. Khairallah, Alexander~M. Rubenchik,
  Philip~J. Depond, and Wayne~E. King.
\newblock {Denudation of metal powder layers in laser powder bed fusion
  processes}.
\newblock {\em Acta Materialia}, 2016.

\bibitem{Wei2015InfluenceComponents}
Kaiwen Wei, Zemin Wang, and Xiaoyan Zeng.
\newblock {Influence of element vaporization on formability, composition,
  microstructure, and mechanical performance of the selective laser melted
  Mg-Zn-Zr components}.
\newblock {\em Materials Letters}, 156:187--190, 2015.

\bibitem{Zhang2020ElementMelting}
Guohao Zhang, Jing Chen, Min Zheng, Zhenyu Yan, Xufei Lu, Xin Lin, and Weidong
  Huang.
\newblock {Element vaporization of Ti-6AL-4V alloy during selective laser
  melting}.
\newblock {\em Metals}, 10(4):1--14, 2020.

\bibitem{Katayama2011DevelopmentVacuum}
Seiji Katayama, Abe Yohei, Masami Mizutani, and Yousuke Kawahito.
\newblock {Development of deep penetration welding technology with high
  brightness laser under vacuum}.
\newblock In {\em Physics Procedia}, 2011.

\bibitem{Jiang2017EffectWelding}
Meng Jiang, Wang Tao, Shuliang Wang, Liqun Li, and Yanbin Chen.
\newblock {Effect of ambient pressure on interaction between laser radiation
  and plasma plume in fiber laser welding}.
\newblock {\em Vacuum}, 2017.

\bibitem{Li2018ExperimentalPressure}
Liqun Li, Genchen Peng, Jiandong Wang, Jianfeng Gong, and Huizhi Li.
\newblock {Experimental study on weld formation of Inconel 718 with fiber laser
  welding under reduced ambient pressure}.
\newblock {\em Vacuum}, 2018.

\bibitem{Jiang2019ComparisonSub-atmosphere}
Meng Jiang, Wang Tao, Yanbin Chen, and Fukang Li.
\newblock {Comparison of processing window in full penetration laser welding of
  thick high-strength steel under atmosphere and sub-atmosphere}.
\newblock {\em Optics and Laser Technology}, 109(August 2018):449--455, 2019.

\bibitem{Sokolov2015ReducedSteel}
Mikhail Sokolov, Antti Salminen, Seiji Katayama, and Yousuke Kawahito.
\newblock {Reduced pressure laser welding of thick section structural steel}.
\newblock {\em Journal of Materials Processing Technology}, 219:278--285, 2015.

\bibitem{Pang20153DEffect}
Shengyong Pang, Xin Chen, Jianxin Zhou, Xinyu Shao, and Chunming Wang.
\newblock {3D transient multiphase model for keyhole, vapor plume, and weld
  pool dynamics in laser welding including the ambient pressure effect}.
\newblock {\em Optics and Lasers in Engineering}, 74:47--58, 2015.

\bibitem{Masmoudi2015InvestigationProcess}
Amal Masmoudi, Rodolphe Bolot, and Christian Coddet.
\newblock {Investigation of the laser–powder–atmosphere interaction zone
  during the selective laser melting process}.
\newblock {\em Journal of Materials Processing Technology}, 225:122--132, 11
  2015.

\bibitem{Bidare2017AnMeasurements}
P.~Bidare, R.R.J. Maier, R.J. Beck, J.D. Shephard, and A.J. Moore.
\newblock {An open-architecture metal powder bed fusion system for in-situ
  process measurements}.
\newblock {\em Additive Manufacturing}, 16:177--185, 8 2017.

\bibitem{Bidare2018LaserPressures}
P.~Bidare, I.~Bitharas, R.M.~M. Ward, M.M.~M. Attallah, and A.J.~J. Moore.
\newblock {Laser powder bed fusion at sub-atmospheric pressures}.
\newblock {\em International Journal of Machine Tools and Manufacture},
  130-131:65--72, 8 2018.

\bibitem{Bidare2018LaserAtmospheres}
P.~Bidare, I.~Bitharas, R.~M. Ward, M.~M. Attallah, and A.~J. Moore.
\newblock {Laser powder bed fusion in high-pressure atmospheres}.
\newblock {\em The International Journal of Advanced Manufacturing Technology},
  pages 1--13, 8 2018.

\bibitem{Zhang2020SimulationFusion}
Xiaobing Zhang, Bo~Cheng, and Charles Tuffile.
\newblock {Simulation study of the spatter removal process and optimization
  design of gas flow system in laser powder bed fusion}.
\newblock {\em Additive Manufacturing}, 32(January 2020):101049, 2020.

\bibitem{Ferrar2012GasPerformance}
B.~Ferrar, L.~Mullen, E.~Jones, R.~Stamp, and C.~J. Sutcliffe.
\newblock {Gas flow effects on selective laser melting (SLM) manufacturing
  performance}.
\newblock {\em Journal of Materials Processing Technology}, 212(2):355--364,
  2012.

\bibitem{Philo2018AProcess}
A.~M. Philo, D.~Butcher, Stuart Sillars, C.~J. Sutcliffe, J.~Sienz, S.~G.R.
  Brown, and N.~P. Lavery.
\newblock {A multiphase CFD model for the prediction of particulate
  accumulation in a laser powder bed fusion process}.
\newblock {\em Minerals, Metals and Materials Series}, Part F3:65--76, 2018.

\bibitem{Reijonen2020OnManufacturing}
Joni Reijonen, Alejandro Revuelta, Tuomas Riipinen, Kimmo Ruusuvuori, and Pasi
  Puukko.
\newblock {On the effect of shielding gas flow on porosity and melt pool
  geometry in laser powder bed fusion additive manufacturing}.
\newblock {\em Additive Manufacturing}, 32(December 2019):101030, 2020.

\bibitem{Shen2020InfluenceFusion}
Haopeng Shen, Paul Rometsch, Xinhua Wu, and Aijun Huang.
\newblock {Influence of Gas Flow Speed on Laser Plume Attenuation and Powder
  Bed Particle Pickup in Laser Powder Bed Fusion}.
\newblock {\em Jom}, 72(3):1039--1051, 2020.

\bibitem{Saunders2017GoneLinkedIn}
Marc Saunders.
\newblock {Gone with the wind - how gas flow governs LPBF performance |
  LinkedIn}, 2017.

\bibitem{Ladewig2016InfluenceProcess}
Alexander Ladewig, Georg Schlick, Maximilian Fisser, Volker Schulze, and Uwe
  Glatzel.
\newblock {Influence of the shielding gas flow on the removal of process
  by-products in the selective laser melting process}.
\newblock {\em Additive Manufacturing}, 10:1--9, 2016.

\bibitem{Anwar2017SelectiveStrength}
Ahmad~Bin Anwar and Quang~Cuong Pham.
\newblock {Selective laser melting of AlSi10Mg: Effects of scan direction, part
  placement and inert gas flow velocity on tensile strength}.
\newblock {\em Journal of Materials Processing Technology}, 240:388--396, 2017.

\bibitem{Kalman2005PickupParticles}
Haim Kalman, Andrei Satran, Dikla Meir, and Evgeny Rabinovich.
\newblock {Pickup (critical) velocity of particles}.
\newblock {\em Powder Technology}, 160(2):103--113, 2005.

\bibitem{Chen2018OptimizationChamber}
Yu~Chen, Guglielmo Vastola, and Yong~Wei Zhang.
\newblock {Optimization of Inert Gas Flow Inside Laser Powder Bed Fusion
  Chamber}.
\newblock In {\em Proceedings of the 29th Annual International Solid Freeform
  Fabrication Symposium – An Additive Manufacturing Conference}, pages
  1931--1939, Austin, TX, USA, 2018.

\bibitem{Buehler2013BuehlerAnalysis}
{Buehler}.
\newblock {\em {Buehler SumMet - A Guide to Materials Preparation and
  Analysis}}.
\newblock Buehler, 4 edition, 2013.

\bibitem{Hann2011AParameters}
D.~B. Hann, J.~Iammi, and J.~Folkes.
\newblock {A simple methodology for predicting laser-weld properties from
  material and laser parameters}.
\newblock {\em Journal of Physics D: Applied Physics}, 44(44), 2011.

\bibitem{Pichler2020MeasurementsSteel}
Peter Pichler, Brian~J. Simonds, Jeffrey~W. Sowards, and Gernot Pottlacher.
\newblock {Measurements of thermophysical properties of solid and liquid NIST
  SRM 316L stainless steel}.
\newblock {\em Journal of Materials Science}, 55(9):4081--4093, 2020.

\bibitem{Kim1975ThermophysicalSteels}
Choong~S. Kim.
\newblock {Thermophysical Properties of Stainless Steels}.
\newblock Technical report, Argonne National Laboratory, Argonne, IL, 1975.

\bibitem{Eagar1983TemperatureSources.}
T.~W. Eagar and N.~S. Tsai.
\newblock {Temperature Fields Produced By Traveling Distributed Heat Sources.}
\newblock {\em Welding Journal}, 62(12):346--355, 1983.

\bibitem{King2014ObservationManufacturing}
Wayne~E. King, Holly~D. Barth, Victor~M. Castillo, Gilbert~F. Gallegos, John~W.
  Gibbs, Douglas~E. Hahn, Chandrika Kamath, and Alexander~M. Rubenchik.
\newblock {Observation of keyhole-mode laser melting in laser powder-bed fusion
  additive manufacturing}.
\newblock {\em Journal of Materials Processing Technology}, 214(12):2915--2925,
  2014.

\end{thebibliography}
\end{document}